\documentstyle[preprint,aps]{revtex}
\begin{document}
%\draft

\title{
{\bf QCD EVOLUTION EQUATIONS FOR HIGH ENERGY PARTONS IN NUCLEAR MATTER}
}

\author{
{\bf K. Geiger}
\footnote{Present address: CERN TH-Division, CH-1211 Geneva 23, Switzerland}
\\
{\it Institute for Theoretical Physics, University of California, Santa Barbara, CA 93106}
\\
{\it School of Physics and Astronomy, University of Minnesota, Minneapolis, MN 55455}
\\ and \\
{\bf B. M\"uller}\\
{\it Department of Physics, Duke University, Durham, NC 27708-0305}
}

\date{\today}

\maketitle

\begin{abstract}
We derive a generalized form of Altarelli-Parisi equations to decribe the
time evolution of parton distributions
in a nuclear medium. In the framework of the
leading logarithmic
approximation, we obtain a set of coupled integro-differential
equations for the parton distribution functions and equations for the 
virtuality (``age'') distribution of partons. 
In addition to parton branching processes, we take 
into account fusion and scattering processes that are specific to QCD in medium.
Detailed balance between gain and loss terms in the resulting evolution equations 
correctly accounts for both real and virtual contributions which yields a 
natural cancellation of infrared divergences. 
\end{abstract}
\noindent

\vspace{0.5cm}

\leftline{PACS Indices: 25.75.+r, 12.38.Bx, 12.38.Mh, 24.85.+p}
\vspace{0.5cm}

\vfill \eject

\noindent {\bf 1. INTRODUCTION}
\medskip

The future
ultra-relativistic heavy ion collider experiments at
the BNL Relativistic Heavy Ion Collider (RHIC)  and
the CERN Large Hadron Collider (LHC) 
are expected to exhibit new phenomena associated with ``QCD in medium'',
i.e. with the microscopic
dynamics of quarks and gluons in the hot, ultra-dense environment
that may be created in the central collision region of these reactions 
\cite{qm,qgp}.  Recently, considerable progress has been made in a better 
understanding of the space-time structure of parton interactions during 
the early stage of these reactions \cite{lbl93,GM92,pcm0}. The conclusion 
emerging from different independent investigations 
\cite{pcmtc,biro,shu93,xnwke}, is that - for RHIC energies and 
beyond---most of the entropy and transverse energy is presumably produced 
already during very early times (within the first 2 fm after the nuclear 
contact) by frequent, mostly inelastic, semihard parton collisions 
involving typical momentum transfers of only a few GeV.

The underlying notion is that the early stage of nuclear collisions at 
sufficiently high energies can well be described in terms of the 
space-time evolution of many internetted parton cascades \cite{qm93}, 
based on renormalization group improved perturbative QCD \cite{muellbook} 
and relativistic kinetic theory \cite{degroot}.  This physical picture 
is motivated by the successful ``semihard QCD'' description of high energy 
hadronic interactions \cite{glr}. The application to ultra-relativistic
nuclear collisions, in particular heavy ion reactions 
\cite{hwakaj,mulblai,mulqm}, assumes that the colliding nuclei may be 
viewed as two coherent clouds of space-like partons with small 
virtualities that materialize into ``real'' excitations due to primary 
parton-parton scatterings.  This primary parton production 
is expected to result in a large initial particle and energy density in the
central collision region, which increases further by subsequent intense
gluon bremsstrahlung, secondary scatterings and rescatterings. 

At some point however, when the parton density becomes so large that 
the quanta begin to overlap in phase-space, recombination (fusion) 
processes become relevant and the density must saturate towards
its limiting value.  It is realized that these semihard processes play 
the major role for the nuclear dynamics at collider energies.  The 
copiously produced minijets \cite{minijet} cannot be considered as 
isolated rare events, but are embedded in complicated multiple 
cascade-type processes, as has been discussed more recently in a number 
of works \cite{pcmtc,pcm0,xnwgyu}. At the same time it is found
\cite{bimuwa,selikov} that color correlations among the initial partons 
randomize so rapidly as the beam particles interpenetrate, such that the 
long range color field effectively vanishes on a space-time scale of a 
small fraction of a fm. Thus the short range character of the 
interactions implies that perturbative QCD can and must be used, and 
that for example the string picture does not apply anymore. 

However, one is still far from a complete and detailed picture, as is 
reflected by the considerable theoretical uncertainty in perturbative 
QCD predictions for global observables  in nucleus-nucleus ($AA$) 
collisions at collider energies, such as particle multiplicities and 
transverse energy production.  The inability to extrapolate accurately 
from of $pp$ ($p\bar p$) data to heavy ion $AA$ collisions is due to 
the current lack of better knowledge about the details of important 
nuclear and dense medium effects.  It is neither surprising, nor 
satisfying, that numerical simulations with QCD based Monte Carlo models
such as HIJING \cite{hijing}, DTUJET \cite{dtujet} or the Parton 
Cascade Model (PCM) \cite{pcm0,pcmapp}, agree very well in describing 
$pp$ collisions at collider energies, but differ in their predictions, 
for e.g. charged particle multiplicities, in heavy ion $AA$ collisions 
by a factor of 2 or more.

The central question is therefore: how is ``QCD in medium'' modified as 
compared to ``QCD in vaccuum''?  In trying to gain a more quantitative 
knowledge about the microscopic parton dynamics in medium, the most 
urgent questions concern (i) the initial conditions regarding the parton 
substructure of large nuclei, in particular the small $x$ region and the
magnitude of nuclear shadowing effects; (ii) the role of color screening 
and color diffusion; (iii) the impact of the Landau-Pomeranchuk-Migdal 
effect; (iv) the space-time dependence of parton interactions with regard 
to the influence of the characteristic interaction times of parton 
scatterings and the formation times for gluons emitted in bremsstrahlung
processes.

Understanding ``QCD in medium'' is also one of the most interesting 
experimental challenges for RHIC and LHC. We hope that this can be 
achieved by analyzing the characteristic space-time structure of parton 
interactions during the very early stage in $pA$ and $AA$ collisions, 
e.g.  by  measuring the production of particles emerging from these 
early times, such as dileptons, direct photons, strange and charmed particles.
\smallskip

Recently it was pointed out by McLerran and Venugopalan \cite{MLV93}
that a consistent perturbative calculation of parton structure functions
at small values of the Bjorken variable $x$ becomes possible when one
considers the limit of a very thick nuclear target. A projectile
propagating at high energy through such a target ``sees'' a very large
area density of valence quarks, and hence experiences an effective
screening of color interactions in the transverse direction. The
condition for the applicability of perturbative QCD then is that the
screening distance is much shorter than the confinement scale
$\Lambda^{-1}$. This is satisfied for a sufficiently thick target
at sufficiently small $x$.

In this paper we want to make use of this insight to explore the
evolution of a parton cascade inside nuclear matter under conditions
where perturbative QCD applies because medium induced effects, such as
color screening and rescattering, provide dynamical cut-offs on a scale
short compared to $\Lambda^{-1}$. In principle, our approach applies
to the propagation of fast partons in any kind of dense medium, be it a
thermalized quark-gluon plasma or ground state nuclear matter. The
equations derived here can therefore be applied to jet quenching in a
QCD plasma \cite{GW91} as well as to the fragmentation cascade of a
quark after deep-inelastic scattering in an infinitely large 
nucleus \cite{frankstrik}.
To keep the discussion specific, we will consider the following
idealized problem. Beginning with a prescribed initial distribution
of fast partons injected into infinitely extended nuclear matter by some
highly localized process of space-time extent, $(Q_0^2)^{-1/2} \ll 
\Lambda^{-1}$, we want to follow the evolution of the
parton distribution in laboratory time, as it propagates through the
medium.

Our plan is to introduce a simplified version of the parton cascade 
approach of Ref. \cite{GM92,pcm0}, which we propose as a supplementary 
diagnostic tool for a more transparent analysis of the aforementioned 
aspects of QCD in medium.  We will reduce the complex space-time 
structure of multiple connected parton cascades to the problem of the 
diffusion of quarks and gluons in dense nuclear matter, for which  we 
can apply an analytical treatment using methods. Our rationale is to 
ignore all processes irrelevant in the present context;  therefore  we
will neglect here the quantitatively important effects of color 
screening, long range color correlations, parton shadowing, etc.
Those aspects may be incorporated in a future extension of this work.

Let us describe in some more detail the essence of this work.  We 
attempt to etablish a connection between the semiclassical probabilistic 
picture of parton evolution in the leading logarithmic approximation (LLA) 
\cite{lla1,lla2,lla3} and the time development of parton cascades in 
six-dimensional phase space within the framework of non-equilibrium
kinetic theory \cite{GM92,pcm0}.  To do so, we need to clarify two 
fundamental issues: First we need to relate the Altarelli-Parisi-Lipatov 
(APL) evolution equations \cite{ap,lipatov}, which determine the change 
of the parton number densities under variation of the variables rapidity 
$y$ and transverse momentum $k_\perp$, or  $x \approx \exp(y)$ and $Q^2 
\approx k_\perp^2$, with the Boltzmann equation, which controls the time
evolution of the phase space densities in both momentum and coordinate space.
Second, we must relate the experimentally accessible number densities $q_i$, 
$\bar q_i$, and $g$ (quarks and antiquarks of flavor $i$, and gluons) 
with the single-particle phase space densities $F_{q_i}$, $F_{\bar q_i}$, 
and $F_g$, respectively.

The first point - the connection between the APL evolution equations and 
the Boltzmann equation---has been previously suggested, tentatively by 
Durand and Putikka \cite{durand}, and explicitly by Collins and Qiu 
\cite{collins}, however, in the formal context of hadron structure which 
is rather different from our considerations of parton cascades in nuclear 
matter.  Although both these authors succeeded in a rederivation of the 
APL equations on the basis of a complete probabilistic picture, they 
stopped short in actually establishing the correspondence between
QCD evolution and the space-time development of the parton distributions.
Nevertheless, the alternative approach of Refs. \cite{durand,collins}
showed that the APL equations can be derived in a statistical manner 
similar to the Boltzmann equation, by taking into account both the gain 
{\it and} the loss of partons due to successive $1\rightarrow 2$ branchings 
in $(x,Q^2)$-space. This self-contained detailed balance eliminates the 
necessity of calculating vertex and wave function renormalization
explicitly, because the loss terms naturally take over this role. As a 
consequence, the resulting evolution equations are free of divergences 
and satisfy the constraints imposed by momentum and quark number 
conservation automatically.

Concerning the second point, we note that the measured parton number densities 
$a_N(x,Q^2)$, where $a \equiv q_i, \bar q_i, g$, give the probability for 
finding  a quark, antiquark, or gluon, inside a nucleon with  fraction 
$x = k_z/P$ of the longitudinal nucleon momentum $P$ and with 
virtuality $Q^2$, or transverse 
momentum $k_\perp^2$. Here $Q^2 \approx k_\perp^2$ sets the scale of hardness
that is identified with the momentum transfer of an interaction of the parton 
with a weakly interacting probe (e.g. a virtual photon) that measures 
the nucleon substructure.  At present, the parton number densities are 
experimentally accessible only in a space-time integrated way and 
therefore must be interpreted as instantanous distributions of partons 
inside a nucleon.  This interpretation, however, has a justified 
operational meaning only in a reference frame in which the nucleon moves 
close to the speed of light, because in such a frame time dilation slows 
the internal motion of the partons such that the statistical picture can 
be applied \cite{kogut}.  On the other hand, in statistical many-particle 
systems the phase-space distribution $F_a(E, \vec k; \vec r, t)$ is the 
probability density for finding a parton of species $a$ in a phase-space 
element $d^3k d^3r$ at time $t$. Evidently $F_a$ contains explicit 
additional information about the space-time structure of the initial 
state nucleons or nuclei, which is only present in an averaged way in 
the measured parton number densities.

On the basis of this knowledge, we will here extend the probabilistic 
approach to the space-time evolution of partons in nuclear matter. To do 
so, we will first relate the $Q^2$-evolution to the development with time $t$, 
and second, we will include not only the $1\rightarrow 2$ branching 
processes, but also the reverse $2\rightarrow 1$ fusion processes and 
in addition $2\rightarrow 2$ scattering processes. The latter two types 
of processes indirectly also give rise to additional stimulated branchings.
Stimulated emission, fusion and scattering processes are naturally absent 
in vacuum, but in medium they are indispensible ingredients for obtaining 
a complete set of of transition amplitudes and a self-consistent parton 
evolution.

The remainder of this paper is organized as follows. In Sec. 2 we set up 
the physical scenario of parton cascade evolution inside nuclear matter 
and introduce the framework of kinetic decription including a consistent
treatment of off-shell propagation of partons. We will relate the quark 
and gluon number densities, as measured by the structure functions, with 
the time-dependent phase space distributions of off-shell partons and will 
translate the QCD evolution of the parton number densities into the 
space-time development of parton cascades.  On the basis of this 
connection, the rates for branching, fusion and scattering processes of 
the partons interacting with the medium are derived. In Sec. 3 we 
generalize the considerations and obtain a set of coupled equations for 
the space-time evolution of quarks, antiquarks and gluons.
Finally, we summarize our results and give some possible future
perspectives. 
\bigskip

\noindent {\bf 2. DESCRIPTION OF PARTON SHOWERS IN NUCLEAR MATTER}
\medskip

\noindent {\bf 2.1 Physical picture of parton showers in medium}
\smallskip

The physical situation that we have in mind is illustrated in Fig. 1.
We consider a parton cascade initiated by a high energy time-like
quark or gluon that has been produced at some point of time $t_0$ 
inside a heavy nucleus due to an external interaction. For instance 
by a collision of a proton with the nucleus, in which case the parton 
is produced by a scattering with another parton of the incoming proton and
is provided with a time-like virtuality $Q^2 \approx p_\perp^2$ where 
$p_\perp^2$ is determined by the momentum transfer in the scattering.
Or by deep inelastic scattering, where the parton is struck by a 
space-like virtual photon with invariant mass squared $q^2 < 0$ and
it acquires a time-like virtuality $Q^2 \approx \vert q^2 \vert$.
In any case, this {\it primary} parton then propagates through the 
partonic matter of the nucleus and initiates of  a  shower of {\it 
secondary} partons.  The attractive feature of such a scenario is that 
it provides a good control of the initial conditions:  the primary 
parton is produced with a well defined four-momentum, determined by the 
momentum transfer of the triggering interaction with the external 
particle. This is to be contrasted with a nucleus-nucleus collision, in 
which case there are many comoving nucleons and the initial state 
contains a rather complicated mixture of initially produced partons with a
a rather broad momentum distribution.

To set a definite physical situation we will from now on consider a 
proton-nucleus ($pA$) collision \cite{pcmpa}.  In order to apply the 
parton picture one has  to go into a frame where both the projectile 
proton and the target nucleus are moving very fast, so that both the 
proton and the nucleons in the nucleus can be resolved into individual 
partons \cite{papi}.  The description of a nucleon as an instantaneous 
distribution of partons at any time requires probing the nucleon over 
time durations and spatial distances small on the scale of internal 
motions of the partons. This condition is fulfilled in any frame of 
reference in which the nucleon moves almost with the speed of light,
because the time dilation effect slows the internal motions such that
the nucleon can be described as a simple quantum mechanical ensemble
of quasireal partons that do not mix with vacuum fluctuations (except 
for the slowest gluons and sea quarks).  It is convenient to choose 
the {\it nucleon-nucleon center-of-mass frame} (CM$_{NN}$)  in which 
each nucleon has the same value of longitudinal momentum $P$ (see Fig. 2),
\begin{equation}
P_z^{(p)} \;=\; + \, P\;\;,\;\;\;\;\;\;\;
P_z^{(A)} \;=\; - \, A\, P
\;\;,\;\;\;\;\;\;\;
\vec P_\perp^{(p)} = \vec P_\perp^{(A)} = \vec 0
\;\;,
\end{equation}
so that 
\begin{equation}
\sqrt{s}\;=\; \sqrt{4 A P^2 \;+\; M_N^2(1 + A^2)}
\end{equation}
and
\begin{equation}
\sqrt{s_{NN}}\;=\; 2\;P
\;\;,
\end{equation}
where $A$ is the nuclear mass number, $M_N$ the nucleon mass, and 
$P/M_N \gg 1$ is assumed, a requirement which is certainly satisfied at 
the colliding beam accelerators RHIC and LHC. For example, at RHIC, the 
maximum $P$ is 250 GeV for $p+p$, 125 GeV for $p+{}^{16}O$ and  100 GeV 
for $p+{}^{197}Au$ \cite{rhic}.  At LHC one has generally a factor of 
30 larger energy available \cite{lhc}.

We assume that the primary quark or gluon originates from the proton 
structure function, carrying longitudinal momentum $k_{z 0} = x_0 P$, and
is provided with an initial time-like virtuality $Q_0^2 > 0$ by a hard 
scattering with one of the nuclear partons.  Hence, this primary parton 
is so to say injected into the nucleus and probes the nuclear environment. 
It will subsequently produce a number of secondary partons by various 
interactions with the nuclear background matter: The incident primary 
parton can either (i) radiate bremsstrahlung gluons,  (ii) collide with partons 
of the nuclear background matter,  or, (iii) absorb these nuclear partons.
The so produced secondary partons will subsequently undergo the same 
type of interactions.  That is, they themselves will lose energy-momentum 
by either exciting partons bound in the nucleus off which they scatter on 
their way, or by radiating gluon bremstrahlung, or (in case of gluons) 
materialize through quark-antiquark pair production. At every new step
the number of particles increases and their average energy and 
longitudinal momentum decreases, until the growing density enhances 
reverse absorption (fusion) processes that may yield a detailed balance, 
or until eventually all the energy-momentum of the primary particle can 
be considered as completely dissipated.  We will refer to this event as 
a {\it parton shower} or {\it parton cascade}.  We emphasize that we 
will explicitly distinguish between the {\it shower partons} on the one 
hand, and the {\it nuclear partons} on the other hand, which are 
coherently bound in the wave function of the nucleus.  Furthermore, it 
should be clear from the above selection of processes (i)-(iii) that  we 
consider here only {\it interactions between the shower partons and the 
nuclear partons}, and not among the shower partons or the nuclear partons 
themselves. In other words, we  consider here the evolution of a single 
cascade and therefore only account for interactions of the cascade with 
the medium but neglect interactions between possible simultaneous
cascades (we will briefly consider a self-interacting cascade at the
end of Section 3).

We will describe the longitudinal evolution of the parton shower along 
the {\it shower axis} ($z$-axis), which we define parallel to the 
direction of momentum of the initiating primary parton.  It is 
convenient to parametrize the four-momenta $k\equiv k^\mu = (E, k_z, 
\vec k_\perp)$ of the shower partons such that for the primary parton,
\begin{equation}
k_0\;=\;\left(x_0 P+\frac{Q_0^2}{2 x_0 P}; \;x_0 P, \;\vec 0 \,\right)
\;\;,
\label{p0}
\end{equation}
whereas for the $j^{th}$ secondary parton,
\begin{equation}
k_j\;=\;\left(x_j P+\frac{Q_j^2+k_{\perp j}^2}{2 x_j P}; \;x_j P, \;
\vec k_{\perp j} \right) \;\;,
\label{pj}
\end{equation}
where $k_{\perp j}^2 < Q_j^2 \ll Q_0^2 \ll P^2$ is assumed, and all 
rest masses are neglected.  It is important to realize that energy and 
momentum are independent variables, since we are dealing with off-shell 
particles of virtuality $Q^2$ with a continuous mass spectrum.

The evolution of a many parton system is described by the change of  
{\it parton number densities} of quarks $q_i$ and antiquarks $\bar q_i$ 
of flavor $i$, or gluons $g$, which are defined as, 
\begin{equation}
a(x, k_\perp^2, Q^2; t)\;=\; \int_0^{Q^2}  d Q^{\prime\,2} \, 
\frac{d N_a(t)}{d x d k_\perp^2 d Q^{\prime\,2}}
\;\;,\;\;\;\;\;\;(\,a\,\equiv \,q_i, \bar q_i, g\,)
\;\;,
\label{axt}
\end{equation}
where
\begin{equation}
N_a(t)\;=\;\int d^3 r \int \frac{d^3 k}{(2 \pi)^3}\,F_a(E,\vec k; t, \vec r)
\;\equiv\; \int \frac{d^3 k}{(2 \pi)^3}\,f_a(E,\vec k; t)
\end{equation}
is the number of partons of type $a$ present at time $t$ with $F_a$ 
denoting the corresponding phase-space density of partons at $d^3k d^3r$, 
and $f_a$ in the second line representing the spatially integrated 
energy-momentum distribution.  Note that we introduced an explicit time 
dependence in the parton number densities (\ref{axt}), and also treat 
$Q^2$ and $k_\perp^2$ as independent variables, in contrast to the usual
identification $Q^2 \approx k_\perp^2$. Furthermore we stress that $Q^2$ 
and $t$ are actually correlated variables, as we will show.  Therefore 
the parton densities are in fact functions of $Q^2$ as well as $t$, 
and we have to describe their $Q^2$-evolution as well as their time 
development.  On the other hand, the experimentally accessible parton 
densities are related to the functions (\ref{axt}) by
\begin{equation}
a_N(x, Q^2)\;=\; \int_{-\infty}^\infty d t \int_0^\infty d k_\perp^2 \; 
a(x, k_\perp^2, Q^2; t) \;\;.
\label{anxt}
\end{equation}

Consequently the invariant mass spectrum in $Q^2$ measures the degree 
of excitation of the shower particles, which will be related to their 
``age'' in the cascade, whereas the variable $x = k_z/ P$ measures the 
change of the longitudinal momenta of the cascade partons, and the
transverse momentum distribution in $k_\perp^2$ reflects the diffusion 
perpendicular to this axis.  According to our chosen geometry, the 
primary parton has no transverse momentum at all, $k_{\perp 0} = 0$, 
but it is off-mass shell by an initial virtuality $Q_0^2$. As the parton 
shower develops in space-time due to branching, fusion and scattering 
processes, the distribution in $x$ will shift to smaller values, the 
distribution in $k_\perp^2$ will broaden, and the in the average the
virtuality $Q^2$ of 
the secondary partons  will decrease. How fast such a cascade evolves 
will depend on the medium properties, which will be reflected in the 
``age'' distribution of shower particles. The denser the medium the 
slower will be the aging process.  In the following section we will 
derive a quantitative formulation of this evolution; however, as a 
motivation let us already here draw a qualitative picture.  Consider
for the moment the evolution of a cascade in vacuum, i.e. in the absence 
of a surrounding medium, so that scattering and fusion processes will 
not be present. In this case, the cascade evolves solely by successive 
branchings.  In the LLA the evolution of time-like virtualities $Q^2$ 
is subject to the ordering condition
\begin{equation}
Q_0^2 \;\gg\; Q_1^2 \;\gg\;\ldots \;\gg\; Q_j^2 \;\gg \; Q_{j+1}^2 \;
\gg \ldots \; \gg \;\mu_0^2 \;\;,
\label{order}
\end{equation}
where $\mu_0$ sets an invariant mass scale at which the perturbative
description of the branching cascade fails.  This strongly ordered 
decrease of virtualities is valid in the kinematic region where 
$p_{z \,j}^2 = (x_j P)^2 \gg Q_j^2 > k_{\perp j}^2$, as we already 
assumed after eq. (\ref{pj}).  It eliminates complicated quantum-mechanical 
interference effects in successive branchings.  In a certain intermediate 
branching in the cascade $k_{j-1} \rightarrow k_j + k_j^\prime$, the 
four-momentum  $k_{j-1}$ is given by (\ref{pj}) and energy-momentum 
conservation at the vertex uniquely relates the momenta of the daughters 
$k_j$ and $k_j^\prime$.  In particular, conservation of longitudinal 
momentum implies
\begin{equation}
x_{j-1} \;=\; x_j\;+\; x_j^\prime
\;\;,
\label{pzcons}
\end{equation}
transverse momentum conservation requires
\begin{equation}
\vec k_{\perp j} \;=\; z_j\, \vec k_{\perp j-1} \;+\;\vec p_{\perp j}
\;\;,\;\quad 
\vec k_{\perp j} {}^\prime \;=\; (1-z_j)\, \vec k_{\perp j-1} \;-\;
\vec p_{\perp j} \;\;,
\label{ptcons}
\end{equation}
and energy conservation yields
\begin{equation}
p_{\perp j}^2 \;\approx\; z_i(1-z_j)\; Q_{j-1}^2 - (1-z_j) Q_j^2-z_j
Q_j^{\prime\; 2} \;\;,
\label{econs}
\end{equation}
where $z_j = x_j/x_{j-1}$ and  $p_{\perp j}^2$ is the squared  intrinsic 
transverse momentum generated with respect to the $\vec k_{j-1}$ direction. 
For simplicity,  we assumed here a symmetric distribution in azimuthal 
angle.  Hence, both the longitudinal and (on average) the squared 
transverse momentum are additive, and the value of $k_{\perp j}^2$ 
with respect to the shower axis is 
determined by the virtualities and the ratios of longitudinal momenta of 
mother and daughters.  Each branching generates a $p_\perp$-kick, so that
the cascade evolves as a random walk of partons in $k_\perp$-space 
\cite{glr}, and at the same time the partons in the cascade become 
increasingly slower and closer to mass shell.

In the presence of nuclear matter this simple evolution is modified by 
scattering and fusion processes.  However the effect of these interactions 
with the medium can be incorporated in a rather straightforward manner: 
In between scatterings or fusions, the successive branchings still 
determine the evolution of the cascade, but instead of evolving 
undisturbed all the way down to $\mu_0^2$ [c.f. (7)], at each vertex of 
interaction with the medium the branching cascade is terminated prematurely, 
and the interacted parton acquires a new virtuality. This sets a new 
starting point from which the parton continues to branch until the next
scattering or fusion, or until it eventually reaches the minimum 
virtuality $\mu_0^2$.  We will show that this modified evolution can be 
cast in terms of a rejuvenation of the cascade, in the sense that each 
interaction of shower partons with the medium ``resets the clock'' by an 
amount that depends on the hardness of the interaction. We will describe 
this mechanism by introducing the concept of the ``age'' of the cascading 
partons.

At this point we would like to comment on some peculiar kinematic 
properties associated with our choice of the CM$_{NN}$ frame for the 
description of parton cascades.  We adopt the convention that the 
momentum along the beam direction of a  shower parton is $k_z = x P$, 
while that of a nuclear parton in the target nucleus is  $k_z^\prime 
=x^\prime P =  -\vert x^\prime\vert P$ (Fig. 2).  Neglecting nuclear 
shadowing, the initial structure functions of the target nucleus are 
($x^\prime < 0$)
\begin{equation}
F_i^{(A)}(x^\prime, Q^2)\;=\; \frac{A}{\pi R_A^2}\; F_i^{(N)}
(\vert x^\prime\vert , Q^2)\;\;,
\end{equation}
where the index $i$ labels the type of parton,  and the functions 
\begin{equation}
F_i^{(N)}(x, Q^2)= e_i^2 [ x \,q_i(x,Q^2)\,+\,x \,\bar q_i(x, Q^2)]
\end{equation}
denote the nucleon structure functions with
\begin{equation}
\sum_i \int _0^1 d x' \; F_i^{(N)}(x', Q^2)\;=\;1
\;.
\end{equation}
One must keep in mind that all nuclear partons have negative momenta 
along the $z$-axis, and hence their $x^\prime$ values are negative.
In the CM$_{NN}$ frame we have the interesting feature that, although 
the distribution of the parton shower particles is initially (at $t=0$)
\begin{equation}
F_i(x, Q^2)\;=\;x\;\delta(x-x_0)\; \delta(Q^2 - Q_0^2)\;\,\delta_{i
i_0}\quad (x_0\,>\,0)
\;\;,
\end{equation}
if we assume a specific primary parton of type $i_0$ with $x_0$ and $Q_0^2$,
the distribution 
$F_i(x, Q^2,t)$
will, with progressing time, eventually take on non-zero values not only 
in the range $0 < x < x_0$, but
also for {\it negative} values of $x$.
This simply means that, in the chosen frame of reference,
the parton cascade initiated by the projectile parton is stopped, and 
finally swept to the left by the bulk matter of the target. 

There are three different types of processes contributing to the momentum 
degradation of the showering partons. A {\it branching} of a parton results 
in two partons that carry smaller momentum fractions $x_1 = z x$ and 
$x_2 = (1-z) x$.  Note that $x_1$ and $x_2$ have the same sign as $x$, 
i.e. the direction of propagation of partons in the CM$_{NN}$ frame is 
never reversed by branching processes.  A {\it scattering} beteen two 
partons $(x_1,Q_1^2)$ and $(x_2,Q_2^2)$ can occur only between partons 
propagating in opposite directions, i.e. if $ x_1 x_2 < 0$. Neglecting
the transverse momentum, the partonic center-of-mass energy squared is
\begin{equation}
\hat s\;=\; 2 \left( |x_1 x_2 | \,-\, x_1 x_2 \right) \, P^2 \;+\;
Q_1^2 \,\frac{|x_1| + |x_2|}{|x_1|} \;+\;
Q_2^2 \,\frac{|x_1| + |x_2|}{|x_2|}
\;\approx\;
4\,|x_1 x_2|\,P^2
\;\;.
\end{equation}
For partons moving in the {\it same} direction, the contribution 
proportional to $P^2$ vanishes and hence their center-of-mass energy is 
too small to allow application of the parton picture of perturbative QCD 
interactions.  On the other hand, two partons moving in the same 
direction in the CM$_{NN}$ frame may undergo {\it fusion}.  The invariant 
mass of the composed parton is given by the same expression as above, 
except that now $x_1 x_2 > 0$ and hence
\begin{equation}
M^2\;=\; \hat s \;=\; Q_1^2 \;\left( 1\,+\,\frac{x_2}{x_1}\right)
\;+\; Q_2^2 \;\left( 1\,+\,\frac{x_1}{x_2}\right)
\;\;.
\label{m2}
\end{equation}
For partons moving in {\it opposite} directions, the virtuality of the fused
state would be comparable to its energy and momentum in the CM$_{NN}$ frame,
violating the basic assumptions underlying the probabilistic parton picture
($Q^2 \ll (xP)^2$).
If we neglect interactions amongst cascading partons, as we will do in 
the following, we therefore have three types of events:
\itemitem{(i)}
a shower parton $(x,Q^2)$ can branch, with $x x_1, x x_2 > 0$;
\itemitem{(ii)}
a shower parton $(x,Q^2)$ can scatter off a parton from the medium, if $x >0$;
\itemitem{(iii)}
a shower parton $(x,Q^2)$ can fuse with a medium parton, if $x <0$.

The visualization of the multiplication of particles as a parton shower 
developing inside the nuclear matter of the target nucleus that we 
sketched in this section is closely related to the picture of the 
parton cascade evolution that underlies the PCM \cite{pcm0}. However,  
the PCM is much more ambitious and complex, but considerably less 
transparent, as it takes into account all kinds of mutual interactions, 
as well as various nuclear and medium effects.  In the present paper we 
shall instead make a number of simplifications and approximations in 
order to illuminate the essentials of the space-time structure of parton 
evolution in nuclear matter.
\medskip

\noindent {\bf 2.2 Time evolution versus $Q^2$-evolution}
\smallskip

As stated, our next step is to find the connection between the time 
evolution of the parton number densities $a(x, k_\perp^2, Q^2; t)$, eq. 
(\ref{axt}), and the well known $Q^2$-evolution of the experimentally 
observable parton densities in a nucleon $a_N(x,Q^2)$, eq. (\ref{anxt}). 
Originally, the $Q^2$-dependence of the structure functions was 
investigated using the method of operator-product expansion. Later
Altarelli and Parisi \cite{ap}, and independently Lipatov \cite{lipatov}, 
derived a set of integro-differential equations for the $Q^2$-evolution 
in the leading logarithmic approximation (LLA) of QCD.  These equations 
are formulated in momentum space with no reference to the space-time 
structure of the parton evolution.  Altarelli and Parisi determined the
$Q^2$-evolution in momentum space by using ``old fashioned'' perturbation 
theory that involved squared $S$-matrix elements with asymptotic, free 
states, integrated over all space and time up to the infinite future.  
This is reasonable for the evolution {\it in vacuum}, where a parton 
cascade simply develops by successive branchings, undisturbed by external 
fields.  However, our objective is to extend this approach to the evolution 
of such parton showers {\it in medium}, i.e. inside nuclear matter
where, in addition to branchings, the shower particles are likely to
undergo multiple interactions with the nuclear partons, so that the 
integration cannot be extended beyond previous and future interaction points.

To state our goal clearly: we want to derive from elementary principles
a kinetic equation for the time-dependent parton number densities 
(\ref{axt}) that
\itemitem{(i)}
relates the space-time evolution of parton cascades to the $Q^2$-evolution 
of virtualities, i.e. the {\it evolution of off shell particles} in time;
\itemitem{(ii)}
accounts for conventional branching processes (QCD {\it in vacuum}), as 
well as for stimulated branchings, fusions, and scatterings in nuclear 
matter (QCD {\it in medium}); 
\itemitem{(iii)}
correctly balances the {\it gain} and {\it loss} of partons in phase-space 
for each of the branching, fusion and scattering subprocesses.
\smallskip

Having set up the problem, we first need to ask the question, how the 
transition amplitudes or probabilities of intermediate parton states 
change if they are restricted to finite time intervals.  Let us start 
by illustrating with rather general considerations how  the time 
dependence in the LLA parton evolution is connected with the change of
the parton virtualities, without making use of the specific form of the
interaction matrix elements.  We will for the moment consider only the 
branching gluons, since they form the dominant component of the parton 
shower \cite{shu93}.  A detailed derivation, including quarks and 
antiquarks, as well as the effect of fusion processes and scattering 
processes that are characteristic for the medium will be addressed in 
the following subsections.

The issue of time-dependent interaction amplitudes is most conveniently 
addressed by using time-dependent perturbation theory in the interaction 
picture \cite{sakurai}, where the transition amplitudes are given by 
the matrix elements of the time evolution operator.  We start with the 
first order transition amplitude $w^{(0)}(t)$ for a gluon in the initial 
state $\vert i\rangle$ to be scattered into the state $\vert f \rangle$ by 
an external interaction and assume that the final state becomes real,
i.e. gets on mass shell, without further decay [see Fig. 3 a)]:
\begin{equation}
w^{(0)}(t) \;=\; (-i)\;\int_{t_i=0}^t d t_0 \;V_0\; e^{i\omega_{fi} t_0}
\;=\;
\frac{V_0}{\omega_{fi}} \;\left(\frac{}{} 1 \,-\,e^{i\omega_{fi} t}\right)
\;\;,
\label{w0}
\end{equation}
where $\omega_{fi} = E_f - E_i$.
Note that the invariant matrix element $V_0 \equiv \langle f \vert 
\hat V \vert i \rangle$, which causes the production of the state 
$\vert f \rangle$ at time $t_0$, has no explicit dependence on time.

Now we consider the second order process [see Fig. 3 b)], where the 
triggering interaction $V_0$ at $t_0$ produces a certain intermediate 
virtual state $\vert a \rangle$ that subsequently decays (branches) into 
the  2-gluon final state $\vert f \rangle=\vert bc\rangle$ at time 
$t_1$ according to the decay matrix element  $V_1(E_a)$.  The transition
amplitude $w^{(1)}(E_a, t)$ depends therefore on the energy of the 
intermediate state and is given by:
\begin{equation}
w^{(1)}(E_a, t) \;=\; (-i)^2\;\int_{t_i=0}^t d t_1 \;V_1(E_a)\; 
e^{i\omega_{fa} t_1} \;\int_{t_i=0}^{t_1} d t_0 \;V_0\; e^{i\omega_{ai} t_0}
\;\;.
\end{equation}
We are interested in the total transition probability, irrespective of 
the energy of the intermediate virtual state $\vert a \rangle$. Hence 
we need to integrate over the continous spectrum $dE_a$.  In doing so 
we define $\tau:= t_1 - t_0$ to be the lifetime of the virtual state and use
$\tau$ and $t_0$ as variables in the integration:
\begin{eqnarray}
w^{(1)}(t) &=& 
\int d E_a \;w^{(1)}(E_a, t) 
\nonumber \\
&=&
(-i)^2\;\int d E_a \;\int_{0}^t d \tau \;V_1(E_a)\; e^{i\omega_{fa} \tau}
\;\int_{0}^{t-\tau} d t_0 \;V_0\; e^{i\omega_{fi} t_0}
\nonumber \\
&=&
i\;\frac{V_0}{\omega_{fi}}\;\int d E_a \;V_1(E_a)\;\int_{0}^t d \tau \; 
e^{i\omega_{fa} \tau} \;\left(\frac{}{} 1\,-\, e^{i\omega_{fi} (t-\tau)}\right)
\;\;.
\label{w1}
\end{eqnarray}
Note that the last term in brackets does not depend on $E_a$.
Now we divide (\ref{w1}) by (\ref{w0}) to obtain the relative amplitude of
the decay of the virtual state and to get information about its average 
lifetime:
\begin{equation}
R(t)\;=\;\frac{w^{(1)}(t)}{w^{(0)}(t)}\;=\;
i\;\int d E_a \;V_1(E_a)\;\int_{0}^t d \tau \; e^{i\omega_{fa} \tau}
\;\left(\frac{1\,-\, e^{i\omega_{fi} (t-\tau)}}
{1\,-\, e^{i\omega_{fi} t}}\right)
\;\;.
\end{equation}
In the limit $t\rightarrow \infty$ we have $E_f \rightarrow E_i$, or 
$\omega_{fi} \rightarrow 0$, and hence for $\tau \ll t$ (i.e. if the 
virtual state lives a short time compared to the overall observation time):
\begin{equation}
R(t)\;\stackrel{t\rightarrow \infty}{\longrightarrow} 
R \;=\;
i\;\int d E_a \;V_1(E_a)\;\int_{0}^\infty d \tau \; e^{i\omega_{fa} \tau}
\;\;,
\end{equation}
because $\exp(i \omega_{fi} \tau) \rightarrow 1$ in this limit.
In order to analyze the $\tau$-dependence let us define the function 
$\rho(\tau)$ through
\begin{equation}
R \;=\;i\;\int_0^\infty d \tau \;\rho(\tau)
\;\;,
\label{R}
\end{equation}
so that
\begin{equation}
\rho(\tau)\;=\;  \int d \varepsilon \; V_1(E_f+\varepsilon)\; 
e^{-i \varepsilon \tau}
\;\;,
\label{rho}
\end{equation}
where $\varepsilon = - \omega_{fa} = E_a - E_f$ characterizes the 
magnitude of virtuality of the intermediate state.  Let us assume that 
$V_2$ limits the virtuality to a range $\vert\varepsilon \vert\,
\lower3pt\hbox{$\buildrel < \over\sim$}\, \varepsilon_{0}$.  In fact, 
this case corresponds to the ordering of virtualities (\ref{order}) in 
the LLA, which implies for a time-like parton cascade strongly decreasing
virtualities of subsequent intermediate states, where for a given 
intermediate state an upper limit of virtuality is constrained by the 
kinematics (\ref{econs}) at the vertex of production of this virtual state.
For the purpose of lucidity, we make a simple Lorentzian ansatz,
\begin{equation}
V_1(E_a)\;=\;\frac{{\cal V}}{1\;+\;\varepsilon^2/\varepsilon_{0}^2}
\;\;,
\end{equation}
with constant ${\cal V}$. Then we get from (\ref{R}):
\begin{equation}
\rho(\tau)\;=\;  {\cal V} \int_{-\infty}^\infty 
d \varepsilon \; \frac{e^{- i \varepsilon \tau}}{1\;+\;
\varepsilon^2/\varepsilon_0^2} \;=\;
 \pi \;{\cal V} \; \varepsilon_{0}\;
e^{- \,\tau \, \varepsilon_{0}}
\;\;.
\end{equation}
As a result the probability
\begin{equation}
\vert \rho(\tau) \vert^2\;=\; 
\pi^2 \,\varepsilon_{0}^2\;
\vert {\cal V}\vert^2 
\;
e^{-2\,\tau \,\varepsilon_{0}}
\label{rho2}
\end{equation}
expresses that the {\it average} life-time $\tau$ of the virtual 
state $\vert a \rangle$ with virtuality $\varepsilon$ in Fig. 3 b)
is determined by the typical virtuality $\varepsilon_{0}$.
In general it is impossible to assign a life time to a particular 
virtuality $\varepsilon$, because these amplitudes interfere coherently 
in the Fourier integral 
(\ref{rho}).

If we parametrize the four-momenta $k^\mu = (E; k_z, \vec k_\perp)$ 
of the particles $a$, $b$, and $c$ in Fig. 3b) as in (\ref{p0}) and (\ref{pj}),
\begin{eqnarray}
k_a&=& \left( \frac{}{} x_a P + \frac{Q_a^2}{2 x_a P}; \;x_a P, \;
\vec 0 \,\right)
\nonumber \\
k_b&=& \left( \frac{}{} x_b P + \frac{Q_b^2+p_\perp^2}{2 x_b P}; \;
x_b P, \;\vec p_\perp \right)
\nonumber \\
k_c&=& \left( \frac{}{} x_c P + \frac{Q_c^2+p_\perp^2}{2 x_c P}; \;
x_b P, \;- \vec p_\perp \right)
\;\;,
\end{eqnarray}
with $x_a = x_b+x_c$ and $P$ denoting the longitudinal momentum defined 
in (1), and $p_\perp^2$ the relative transverse momentum squared 
generated in the branching, then 
\begin{equation}
\varepsilon_0 \;\equiv\;\varepsilon_a \;=\; \frac{Q_a^2}{2\vert x_a
\vert P}
\;\;,
\end{equation}
sets the upper limit for the virtualities of the daughter partons with 
$Q_b^2$ and $Q_c^2$ determined as in eq. (\ref{econs}).  Hence eq. 
(\ref{rho2}) implies that the transition probability $\vert 
\rho(\tau)\vert^2$ is appreciable only for those final states $\vert f 
\rangle = \vert b c \rangle$ that satisfy
\begin{equation}
\tau \;
\,\lower3pt\hbox{$\buildrel < \over\sim$}\, 
\;\tau_a \;=\; \frac{1}{2\,\varepsilon_{0}}\;=\;\frac{\vert x_a\vert P}{Q_a^2}
\;\;.
\label{tau}
\end{equation}
\medskip

\noindent {\bf 2.3 Derivation of the time-dependence of fragmenting 
parton cascades}
\smallskip

Let us extend these considerations to the evolution of a parton cascade in
the LLA with many intermediate virtual states. Recall the situation that 
we sketched in Sec. 2.1, in which the cascade is initiated by a primary 
parton, say a gluon, with virtuality $Q_0^2$ at time $t_0$. By successive 
gluon emissions (branchings) the cascade evolves with strongly ordered 
decreasing virtualities $Q_i^2 \gg Q_{i+1}^2$ from $Q_0^2$ down to $\mu_0^2$
[c.f. eq. (\ref{order})], so that interference terms can be neglected in 
this approximation. This is illustrated in Fig. 4a). Consequently, each 
branching occurs in the average at a certain time $t_j(Q_j^2)$
with $t_j \ll t_{j+1}$, and the evolution stops at $t_f(\mu_0^2)$.
In analogy to (\ref{tau}) the average life time of the $j^{th}$ gluon 
$g_j$ is given by
\begin{equation}
\tau_j \;= \;
\frac{\vert x_j\vert P}{Q^2}
\;\;,
\end{equation}
Over this time the two gluons $g_j$ and $g_j^\prime$
separate by
\begin{equation}
\Delta r_{\perp j}\;=\; v_{\perp j}\;\tau_j \;=\; 
\frac{k_{\perp j}}{Q_j^2}\;
\,\lower3pt\hbox{$\buildrel < \over\sim$}\, \frac{1}{Q_j}
\;\;,
\end{equation}
where $v_{\perp j} = k_{\perp j}/E_j \approx k_{\perp j}/(x_j P)$ and
and $k_{\perp j}\equiv \vert \vec k_{\perp 1} \vert$.
These expressions are in agreement with the uncertainty principle. 

>From this simple consideration we can conclude that such a parton 
cascade, when embedded inside nuclear matter, develops a transverse 
cross section $\pi r_\perp^2(t)$ that grows linearly with time, as 
illustrated in Fig. 4b). That is, with progressing time the interaction 
with the surrounding medium  becomes increasingly probable, so that 
scattering and fusion processes need to be taken into account. In this 
case, the sequence of successive branchings is terminated naturally when 
the $n^{th}$ gluon collides or fuses with another parton of the medium. 
The gluon is then re-excited by the momentum transfer of the interaction 
and it can subsequently start a new cascade sequence of branchings, and 
the game repeats.
\smallskip 

We proceed now deriving a quantitative formulation of this picture.
At first we will consider the time evolution in $Q^2$ and $x$; the 
lateral shower development with respect to the transverse momentum 
$k_\perp^2$ will be discussed subsequently.  We will study the evolution 
of the gluon longitudinal momentum distribution
\begin{equation}
g(x,t) \;=\; \int dk_\perp^2 \; g(x,k_\perp^2, t)
\;\;,
\label{g0m}
\end{equation}
that is the zeroth moment in $k_\perp^2$ of the full gluon distribution
$g(x,k_\perp^2,t)$, defined by eq. (\ref{axt}). 
Now let us introduce the integrated {\it time-like Sudakov factor}, 
or non-branching probability,
\begin{equation}
T(t)\;\equiv\; T(Q_0^2, \mu_0^2; t_0, t)
\;\;,
\end{equation}
which is the probability that {\it no} branching occurs whatsoever 
between $Q_0^2$ and $\mu_0^2$ within the time span between $t_0$ and 
$t_0+t$. We will set the clock at $t_0=0$, so that the time-dependence 
resides solely in $t$.  In a diagrammatic analysis of the branching
cascade, the Sudakov factor arises from loop diagrams which restore
unitarity \cite{collins}.  Note that the Sudakov factor has the 
propagator property $T(Q_0^2,\mu_0^2;t_0,t)=T(Q_0^2, Q_1^2; t_0, t_1) 
T(Q_1^2,\mu_0^2; t_1, t)$, or short $T(t)=T(t_1) T(t-t_1)$, of which 
we will make use.

The probability that a branching of a parton actually does occur 
between $Q^2$ and $Q^2+dQ^2$, and between $t$ and $t+dt$ is given by
\begin{equation}
\Psi(z,Q^2;t) \, dz\, dQ^2\, dt\;=\; \psi_Q(z,Q^2) dz\,dQ^2\;\,\psi_t(t) \,dt
\;\;,
\end{equation}
Here the distribution in $Q^2$,
\begin{equation}
\psi_Q(z,Q^2) \,d z\,d Q^2\;=\; \frac{\alpha_s(Q^2)}{2 \pi Q^2}\,dQ^2\;
\gamma_{a\rightarrow bc}(z) \, dz
\end{equation}
is the usual Altarelli-Parisi branching kernel \cite{ap} in which 
$\gamma_{a\rightarrow bc}(z)$ is the longitudinal momentum distribution
of the two daughter partons $b$ and $c$, characteristic for the type of 
branching,
\begin{eqnarray}
\gamma_{q \rightarrow q g} (z) &=&
\frac{4}{3} \,\left( \frac{1 + z^2}{1 - z} \right)
\nonumber
\\
\gamma_{g \rightarrow g g} (z) &=&
6 \,\left( \frac{z}{1-z} + \frac{1-z}{z} +  z (1 - z) \right)
\nonumber
\\
\gamma_{g \rightarrow q \bar q} (z) &=&
\frac{1}{2} \, \left( z^2 + (1 - z)^2 \right)
\;\;,
\label{gamma}
\end{eqnarray}
and
the running QCD coupling strength in one-loop order is as usual 
\begin{equation}
\alpha_s (Q^2)\;=\;
\frac{12 \pi}{(33-2 f) \ln (Q^2/\Lambda^2)}
\;\;\;,
\end{equation}
where $\Lambda$ is the QCD renormalization scale 
and $f$ is the number of quark
flavors that can be probed at scale $Q^2$. 

The time of branching is distributed according to a normalized distribution
\begin{equation}
\psi_t(t)\,dt\;=\; \frac{Q^2}{xP}\;f\left(\frac{Q^2}{xP} \,t\right)\;dt
\;\;,
\end{equation}
where 
$Q^2/(xP)$ represents 
the life time of the gluon due to its virtuality in the laboratory frame, 
and the normalization is such that $\int_0^\infty d\xi f(\xi) =1$.
For example, for exponential decay, we have
\begin{equation}
f\left(\frac{Q^2}{xP} \,t \right)\;=\; \exp\left(- \frac{Q^2}{xP} \,t \right)
\;\;.
\end{equation}

We will now derive an evolution equation for the gluon number density 
that correlates the virtualities of the partons in the cascade with the 
times of branching by calculating the Sudakov factor.  In order not to 
burden the discussion unnecessarily in this section, we restrict the 
cascade to containing gluons only, and we solely consider interactions 
with gluons in the nuclear medium. We therefore have to evolve a single 
function, the gluon distribution $g(x,t)$.  To do so, we expand the 
evolution of $g(x,t)$ into an infinite sum of contributions, 
\begin{equation}
g(x,t)\;=\; \sum_{n=0}^\infty \,g^{(n)}(x,t)
\;\;,
\label{sumg}
\end{equation}
represented by Feynman diagrams for $n$ successive branchings at 
$Q_n^2$ and $t_n$, as depicted in Fig. 5. The {\it initial gluon 
distribution} at $Q_0^2$ and $t_0=0$ is denoted as $g_0(x)$.
\smallskip

(i) If no branching occurs between $Q_0^2$ and $\mu_0^2$ during a time $t$,
the only change is that the probability of finding gluons that have not 
branched at all decreases; this is expressed by the Sudakov factor (34),
such that [c.f. Fig. 5]
\begin{equation}
g^{(0)}(x,t) \;=\; T(t)\; g_0(x)
\;\;.
\label{g0}
\end{equation}

(ii) The probability that a single branching occurs somewhere between 
$Q_0^2$ and $\mu_0^2$, within the time interval $t$, is given by the 
probability that no branching occurs between $Q_0^2$ and $Q_1^2$ up to $t_1$ 
[denoted as $T(t_1) = T(Q_0^2,Q_1^2;0,t_1)$],
times the probability for a branching at $Q_1^2$ and $t_1$ [i.e. 
$\Psi(z_1,Q_1^2;t_1)$] convoluted with the gluon distribution $g_0$, 
times the probability that no further branching occurs between $Q_1^2$ 
and $\mu_0^2$, in the time interval between $t_1$ and $t$ 
[denoted as $T(t-t_1)=T(Q_1^2,\mu_0^2;t_1,t)$]. This has to be
integrated over all possible intermediate  $Q_1^2$ and $t_1$ [c.f. Fig. 5]. 
Accordingly we have
\begin{eqnarray}
g^{(1)}(x,t) & = &
\int_0^{t} d t_1\,\int_{\mu_0^2}^{Q_0^2} dQ_1^2 \; T(t_1)\;
\;\int_0^1 \frac{d z_1}{z_1} \;\Psi(z_1,Q_1^2; t_1)\;
T(t-t_1)
\nonumber \\
&=&
T(t)\;\int_{\mu_0^2}^{Q_0^2} dQ_1^2 \
\frac{\alpha_s(Q_1^2)}{2 \pi Q_1^2}
\;\int_0^1 \frac{d z_1}{z_1} \,\gamma_{g\rightarrow gg}(z_1)\; 
g_0\left(\frac{x}{z_1}\right) \;\frac{z_1 Q_1^2}{x P} \,
\int_0^t dt_1\;f\left(\frac{z_1 Q_1^2}{xP}\, t_1\right)
\;\;,
\end{eqnarray}
where $z_1 = x_1/x$. We observe that the effect of the time integral is to 
limit the range of $Q_1^2$-values which contribute to $g^{(1)}$. For large 
values of $Q_1^2$, the time integral yields unity (all branchings
have occurred), whereas for $Q_1^2$ small, the time integral tends to zero.
In order to make progress, we simplify the expression by replacing 
the integrated decay rate 
by a step function:
\begin{equation}
g^{(1)}(x,t) \;=\;
T(t)\;\int_{\mu_0^2}^{Q_0^2} dQ_1^2 \;
\frac{\alpha_s(Q_1^2)}{2 \pi Q_1^2}
\;\int_0^1 \frac{d z_1}{z_1} \,\gamma_{g\rightarrow gg}(z_1)\; 
g_0\left(\frac{x}{z_1}\right) \;\theta\left(\frac{z_1 Q_1^2}{xP} t 
\,-\,1\right)
\;\;.
\end{equation}
This is certainly a crude approximation, but it correctly embodies 
the uncertainty relation between time and virtuality, that is $t \ge \tau_1 
= \gamma_1/ Q_1$, where the Lorentz factor is $\gamma_1 = xP/Q_1$.
Now we differentiate and perform the integral over the virtuality $Q_1^2$:
\begin{eqnarray}
t\,\frac{\partial}{\partial t}\,
g^{(1)}(x,t) & = &
\;t\,\frac{\partial T(t)}{\partial t}\; \frac{g^{(1)}(x,t)}{T(t)}
\nonumber \\
& & \;\;+\;
T(t)\;\int_{\mu_0^2}^{Q_0^2} dQ_1^2 
\frac{\alpha_s(Q_1^2)}{2 \pi}
\;\int_0^1 d z_1 \,\gamma_{g\rightarrow gg}(z_1)\; g_0\left(\frac{x}{z_1}\right)
\;\frac{t}{xP}\,\delta\left(\frac{z_1 Q_1^2}{xP} t -1\right)
\nonumber \\
& = &
\left(t\,\frac{\partial}{\partial t} \ln T(t) \right) \; g^{(1)}(x,t)
\nonumber \\
& & \;\;+\;
\int_0^1 \frac{d z}{z}\; 
\frac{\alpha_s\left(\frac{x P}{z t}\right)}{2 \pi}
\;\theta\left(\frac{x P}{z t} -\mu_0^2\right)
\;\gamma_{g \rightarrow gg}(z) \;
\; g^{(0)}\left(\frac{x}{z},t\right)
\;\;,
\end{eqnarray}
where in the last step we set $z=z_1$ and used eq. (\ref{g0}) to absorb 
the Sudakov factor in front of the second term. We see that
the first order (single-branching) contribution $g^{(1)}$ is determined
by the zeroth order (no-branching) contribution $g^{(0)}$.
\smallskip

(iii) The same arguments hold when we calculate the diagram containing 
$n$ branchings in the LLA. One obtains $g^{(n)}$ as an integral over 
the branching of $g^{(n-1)}$, with the result:
\begin{eqnarray}
t\,\frac{\partial}{\partial t}\,
g^{(n)}(x,t) &=&
\left(t\,\frac{\partial}{\partial t} \ln T(t) \right) \; g^{(n)}(x,t)
\nonumber \\
& & \;\;+\;
\int_0^1 \frac{d z}{z}\; 
\frac{\alpha_s\left(\frac{x P}{z t}\right)}{2 \pi}
\;\theta\left(\frac{x P}{z t} \,-\,\mu_0^2\right)
\;\gamma_{g \rightarrow gg}(z) \;
\; g^{(n-1)}\left(\frac{x}{z},t\right)
\;\;.
\label{gn}
\end{eqnarray}
\smallskip

(iv) According to eq. (\ref{sumg}), the {\it total} gluon distribution 
$g(x,t)$ at time $t$ can now be obtained by summing all contributions 
$g^{(n)}(x,t)$,  yielding the following differential equation:
\begin{eqnarray}
t\,\frac{\partial}{\partial t}\,
g(x,t) 
&=&
\left(t\,\frac{\partial}{\partial t} \ln T(t) \right) \; g(x,t)
\nonumber \\
& & \;\;+\;
\int_0^1 \frac{d z}{z}\; 
\frac{\alpha_s\left(\frac{x P}{z t}\right)}{2 \pi}
\;\theta\left(\frac{x P}{z t} \,-\,\mu_0^2\right)
\;\gamma_{g \rightarrow gg}(z) \;
\; g\left(\frac{x}{z},t\right)
\;\;.
\label{gxt}
\end{eqnarray}
This equation has the same form as the APL equation 
for the gluon distribution, except that $Q^2$ is everywhere replaced by the
variable $\frac{xP}{zt}$ and, since that variable depends on the ratio 
$z = x_1/x$, the coupling $\alpha_s(Q^2)$ cannot be taken outside
of the $z$-integral.

(v) Next we need to calculate the unitary restoring Sudakov factor 
$T(t)$. It describes the loss of probability for having no gluon
branchings at all up to time $t$. Thus we have to accumulate the 
probability for having no branching, one branching, two branchings, etc., 
and then determine $T(t)$ from the condition that the total probability 
remains unity at all times:
\begin{equation}
1\;=\; P(t) \;=\; \sum_{n=0}^{\infty} \, w^{(n)}(t)
\;\;.
\label{pt}
\end{equation}
Here $w^{(n)}(t)$ denotes the probability for having $n$ branchings 
up to time $t$.  The probability for no branching at all is evidently
\begin{equation}
w^{(0)}(t) \;\,=\;\,\frac{\int_0^1 d x \, g^{(0)}(x,t)}{\int_0^1 d x \, 
g_0(x)} \;\,=\;\, T(t)
\;\;,
\label{w0t}
\end{equation}
where $g_0$ in the denominator is the initial gluon distribution at 
$t_0=0$ and $Q_0^2$.  The probability for one single branching is
\begin{equation}
w^{(1)}(t) \;\,=\;\,\frac{1}{2}\;\frac{\int_0^1 d x \, g^{(1)}(x,t)}
{\int_0^1 d x \, g_0(x)} \;\;,
\label{w1t}
\end{equation}
where the factor 1/2 in front arises because $g^{(1)}(x,t)$ counts every 
branching twice, since it counts the number of gluons (two per branching). 
Using (\ref{gn}), we find for 
the integral in the numerator:
\begin{eqnarray}
\int_0^1 dx \;g^{(1)}(x,t) &=&
T(t)\;\int_{\mu_0^2}^{Q_0^2} dQ^2 \;
\frac{\alpha_s(Q^2)}{2 \pi Q^2}
\int_0^1 \frac{d z}{z}\; 
\gamma_{g \rightarrow gg}(z) \;
\int_0^1 d x \;g_0\left(\frac{x}{z}\right)
\;\theta\left(\frac{z Q^2}{x P} t \,-\,1\right)
\nonumber \\
& = &
T(t)\;\int_{\mu_0^2}^{Q_0^2} dQ^2 \;
\frac{\alpha_s(Q^2)}{2 \pi Q^2}
\int_0^1 d z\; 
\gamma_{g \rightarrow gg}(z) \;
\int_0^1 d\xi \;g_0\left(\xi\right)
\;\theta\left(\frac{Q^2}{\xi P} t \,-\,1\right)
\;\;,
\end{eqnarray}
where we have set $\xi = x/z$. Now we again differentiate with respect 
to $t$ and carry out the $Q^2$-integration:
\begin{eqnarray}
t\,\frac{\partial}{\partial t}\,
\left(\int_0^1 d x \;g^{(1)}(x,t)\right) 
&=&
\left(t\,\frac{\partial}{\partial t} \ln T(t) \right) \; 
\int_0^1 d x \;g^{(1)}(x,t)
\nonumber \\
& &\;\;+\;
\int_0^1 d z \gamma_{g \rightarrow gg}(z) 
\;\int_0^1 d \xi \;
\frac{\alpha_s\left(\frac{\xi P}{t}\right)}{2 \pi}\;
\theta\left(\frac{\xi P}{t} \,-\,\mu_0^2\right)
\; g_0\left(\xi\right)
\;T(t)
\;\;.
\end{eqnarray}
Hence, using (\ref{g0}) and (\ref{w0t}) and renaming the integration
variable, we find:
\begin{eqnarray}
t\,\frac{\partial}{\partial t}\,
w^{(1)}(t) 
&=&
\left(t\,\frac{\partial}{\partial t} \ln T(t) \right) \; w^{(1)}(t)
\nonumber \\
& & \;\;+\;
\frac{1}{2}\;\int_0^1 d z \;\gamma_{g \rightarrow gg}(z)\;
\frac{
\;\int_0^1 d x \;
\frac{\alpha_s\left(\frac{x P}{t}\right)}{2 \pi}\;
\theta\left(\frac{x P}{t} \,-\,\mu_0^2\right)
\; g^{(0)}\left(x,t\right)
}{
\int_0^1 d x \,g^{(0)}(x,t)
\; w^{(0)}(t)
}\;\;.
\end{eqnarray}
We can generalize this result easily to the case of $n$ branchings:
\begin{eqnarray}
t\,\frac{\partial}{\partial t}\,
w^{(n)}(t) 
&=&
\left(t\,\frac{\partial}{\partial t} \ln T(t) \right) \; w^{(n)}(t)
\nonumber \\
& & \;\;+\;
\frac{1}{2}\;\int_0^1 d z \;\gamma_{g \rightarrow gg}(z)\;
\frac{
\;\int_0^1 d x \;
\frac{\alpha_s\left(\frac{x P}{t}\right)}{2 \pi}\;
\theta\left(\frac{x P}{t} \,-\,\mu_0^2\right)
\; g^{(n-1)}\left(x,t\right)
}{
\int_0^1 d x g^{(n-1)}(x,t)
}\; w^{(n-1)}(t)
\;\;,
\end{eqnarray}
where the generalization of (\ref{w1t}) is
\begin{equation}
w^{(n)}(t) \;\,=\;\,\frac{1}{2^n}\;\frac{\int_0^1 d x \, 
g^{(n)}(x,t)}{\int_0^1 d x \, g_0(x)}
\;\;.
\end{equation}
This equation can be summed over all $n$, assuming that the 
{\it average} running coupling
\begin{equation}
\overline{\alpha_s}\left(t;\mu_0^2\right)\;\,\equiv\;\,
\frac{
\;\int_0^1 d x \;
\alpha_s\left(\frac{x P}{t}\right)\;
\theta\left(\frac{x P}{t} \,-\,\mu_0^2\right)
\; g^{(n-1)}\left(x,t\right)
}{
\int_0^1 d x \,g^{(n-1)}(x,t)
}
\label{alp1}
\end{equation}
does not depend on the number of branchings. Since $\alpha_s(Q^2)$ is 
a slowly varying function of $Q^2$, this should be a reasonable 
approximation. Then we have with eq. (\ref{pt}),
\begin{equation}
0\;=\;
t\,\frac{\partial}{\partial t}\, P(t)\;=\;
\left( t\,\frac{\partial}{\partial t}\, \ln T(t) \right) \; P(t) \;+\;
\frac{1}{2}\;\int_0^1 d z \;\gamma_{g \rightarrow gg}(z)\;
\;\frac{\overline{\alpha_s}\left(t;\mu_0^2\right)}{2 \pi} \; P(t)
\;\;,
\end{equation}
or
\begin{equation}
t\,\frac{\partial}{\partial t}\, \ln T(t)\;=\;
-\;\frac{1}{2}\;\int_0^1 d z \;\gamma_{g \rightarrow gg}(z)\;
\;\frac{\overline{\alpha_s}\left(t;\mu_0^2\right)}{2 \pi} 
\;\;.
\end{equation}

(vi) This result is now reinserted into the equation (\ref{gxt}) 
for $g(x,t)$, yielding:
\begin{eqnarray}
t\,\frac{\partial}{\partial t}\,
g(x,t) 
&=&
-\;\frac{1}{2}\;\int_0^1 d z \;\gamma_{g \rightarrow gg}(z)\;
\;\frac{\overline{\alpha_s}^g\left(t\right)}{2 \pi} 
\; g(x,t) 
\nonumber \\
& &\;\;+\;
\int_0^1 \frac{d z}{z}\; 
\frac{\alpha_s\left(\frac{x P}{z t}\right)}{2 \pi}
\;\theta\left(\frac{x P}{z t} \,-\,\mu_0^2\right)
\;\gamma_{g \rightarrow gg}(z) \;
\; g\left(\frac{x}{z},t\right)
\;\;,
\label{gxtf}
\end{eqnarray}
where now (\ref{alp1}) is replaced by the average coupling of gluons
\begin{equation}
\overline{\alpha_s}^g\left(t\right)\;\,\equiv\;\,
\frac{
\;\int_0^1 d x \;
\alpha_s\left(\frac{x P}{t}\right)\;
\theta\left(\frac{x P}{t} \,-\,\mu_0^2\right)
\; g\left(x,t\right)
}{
\int_0^1 d x \,g(x,t)
}
\;\;.
\label{alp2}
\end{equation}

We emphasize three important properties of the evolution equation 
(\ref{gxtf}):  First, the probabilistic method used here to derive the 
evolution of the gluon distribution automatically yields both the 
contributions of loss and gain of gluons  that modify $g(x,t)$ [first, 
respectively, second term in (\ref{gxtf})].  We furthermore note that 
for asymptotic times $t\rightarrow \infty$ the evolution equation
recovers the time-independent APL equation for the gluon number density. 
In this limit all the partons have evolved down to virtuality $\mu_0^2$,
so that the total gluon number density depends only on the scale $\mu_0^2$.
The apparent singularity at $t=0$ in the argument of $\alpha_s$ in 
(\ref{gxtf}) and (\ref{alp2}) causes no problem because $t=0$ never 
occurs since the earliest point of time is associated with the very 
first branching of the initial gluons with virtuality $Q_0^2$, which 
can occur only after the finite time $t = xP/Q_0^2 > 0$.
\smallskip

(vii) For cascading quarks and antiquarks, allowing only for the 
process $q \rightarrow q g$ (the full equations will be derived in Sec. 3),
we have a similar equation, except for the lack of the factor 1/2 in the 
loss term:
\begin{eqnarray}
t\,\frac{\partial}{\partial t}\,
q(x,t) 
&=&
-\;\int_0^1 d z \;\gamma_{q \rightarrow qg}(z)\;
\;\frac{\overline{\alpha_s}^q(t)}{2 \pi} 
\; q(x,t) 
\nonumber \\
& & \;\;+\;
\int_0^1 \frac{d z}{z}\; 
\frac{\alpha_s\left(\frac{x P}{z t}\right)}{2 \pi}
\;\theta\left(\frac{x P}{z t} \,-\,\mu_0^2\right)
\;\gamma_{q \rightarrow qg}(z) \;
\; q\left(\frac{x}{z},t\right)
\;\;,
\end{eqnarray}
where in this case the average coupling is determined by the quark density
\begin{equation}
\overline{\alpha_s}^q(t)\;\,\equiv\;\,
\frac{
\;\int_0^1 d x \;
\alpha_s\left(\frac{x P}{t}\right)\;
\theta\left(\frac{x P}{t} \,-\,\mu_0^2\right)
\; q\left(x,t\right)
}{
\int_0^1 d x \,q(x,t)
}
\;\;.
\end{equation}
Note that not only is the branching function $\gamma_{q\rightarrow q g}$ 
different from the gluon branching case, but also $\overline{\alpha_s}$ 
depends implicitly on the quark number density instead of the gluon 
distribution.  The conservation of quark number is derived by integrating 
the equation over $x$:
\begin{eqnarray}
t\,\frac{\partial}{\partial t}\,
\int_0^1 d x\;q(x,t) 
&=&
-\;\int_0^1 d z \;\gamma_{q \rightarrow qg}(z)\;
\;\frac{\overline{\alpha_s}(t)}{2 \pi} 
\; \int_0^1 dx \;q(x,t) 
\nonumber \\
& & \;\;+\;
\int_0^1 \frac{d z}{z}\; 
\int_0^1 d x\; 
\frac{\alpha_s\left(\frac{x P}{z t}\right)}{2 \pi}
\;\theta\left(\frac{x P}{z t} \,-\,\mu_0^2\right)
\;\gamma_{q \rightarrow qg}(z) \;
\; q\left(\frac{x}{z},t\right)
\;\;.
\end{eqnarray}
Introducing the new integration variable $\xi = x/z$ in the second term, one immediately
finds that it equals the first term, but with positive sign, so that
\begin{equation}
t\,\frac{\partial}{\partial t}\,
\int_0^1 d x\;q(x,t) \;=\;0
\;\;.
\end{equation}
\medskip

\noindent {\bf 2.4 Scattering and fusion processes}
\smallskip

As pointed out in Sec. 2.1,
scattering and fusion processes rejuvenate the
fragmentation cascade. Let us first consider elastic scatterings
between two partons. If a parton (gluon) in an evolving cascade
scatters on its way with some other parton from the nuclear medium
involving a transverse momentum exchange $p_\perp^2$, then its maximal
virtuality is reset to $Q^2 = p_\perp^2$. Since we have related the
virtuality $Q^2$ to the evolution of laboratory time $t$ via the
relation
\begin{equation}
t=xP/Q^2
\label{txp}
\end{equation}
we therefore need to ``reset the clock'' for a parton after each
scattering process, if the momentum transfer $p_\perp^2$ 
is larger than the present virtuality of the parton. This is illustrated in Fig. 6.
To keep track of these repeated rejuvenations, we introduce a new
independent variable $\tau$ to denote the {\it age of a parton}. Since
the scatterings and fusion processes in a medium occur stochastically,
the connection between laboratory time $t$ and the age $\tau$
cannot be deterministic, but must rather have the nature of a
probability distribution $A(\tau,t,x)$. Here the distribution $A$ denotes
that a parton with longitudinal momentum fraction $x$ has age $\tau$ at
the laboratory time $t$. As will become clear below, the age
distribution will differ for different values of $x$. 
It is however much easier to derive an evolution equation for the quantity
\begin{equation}
g(x,\tau,t)\;=\; A(\tau,t,x)\; g(x,t)
\end{equation}
rather then for the age distribution itself. The function $g(x,\tau,t)$ describes
the number of gluons that have longitudinal momentum $x P$ and virtuality
$Q^2 = |x|P/\tau$. Of course, if $g(x,\tau,t)$ is known, the age distribution
is easily recovered as
\begin{equation}
A(\tau,t,x)\;=\; \frac{g(x,\tau,t)}{\int_0^\infty d\tau\,g(x,\tau,t)}
\;\;.
\end{equation}

The age distribution evolves in parallel with laboratory time $t$ in
between scatterings and is set back to the age
$\tau(p_\perp^2)=|x|P/p_\perp^2$,
corresponding to the maximal virtuality $p_\perp^2$ acquired in the
collision. The resetting of the clock occurs only, if $\tau(p_\perp^2$) 
is younger than the present age of the parton, because otherwise the
present scattering process cannot be separated from the previous
interaction that kicked the parton off mass shell.  For simplicity,
let us here consider only gluon partons in the cascade and in the medium. 
The time development of $g(x,\tau,t)$ is governed by four different contributions:
\itemitem{(i)}
a ``free-streaming'' term describing that $\tau$ evolves parallel with $t$, if the
parton does not interact, i.e. the parton ``ages naturally'';
\itemitem{(ii)}
a term describing that a branching parton transmits its virtuality to its daughter partons,
i.e. these acquire a new age $z \tau$ according to the branching fraction $z$, so that the variable 
$Q^2 = |x| P/\tau$ remains unchanged;
\itemitem{(iii)}
a term describing the rejuvenation of partons in two-body scattering processes; and
\itemitem{(iv)}
a term describing the change in the virtuality of a parton by fusion with a parton from the 
nuclear medium.

\noindent
Hence,
\begin{equation}
\frac{\partial}{\partial t} \,g(x,\tau,t) \;=\;
\;\left(\frac{\partial g}{\partial t}  \right)_{free} \;+\;
\;\left(\frac{\partial g}{\partial t}  \right)_{branch} \;+\;
\;\left(\frac{\partial g}{\partial t}  \right)_{scatt} \;+\;
\;\left(\frac{\partial g}{\partial t}  \right)_{fus} 
\;\;.
\end{equation}

(i)
The free-streaming term is easily obtained from the condition that $\tau$ and
$t$ evolve in this case parallel, requiring that
\begin{equation}
g(x,\tau,t)\;=\; g(x,\tau+d \tau, t+ dt) \;=\;
g(x,\tau,t)\;+\;
\frac{\partial g}{\partial \tau}  \,dt\;+\;
\frac{\partial g}{\partial t} \,dt
\end{equation}
in the absence of other interactions. 
Hence,
\begin{equation}
\left(\frac{\partial g}{\partial t}\right)_{free} \;=\;
-\;\frac{\partial}{\partial \tau}\,g(x,\tau,t)
\;\;.
\end{equation}

(ii) The branching term is obtained simply from eq. (\ref{gxtf})
after division by $t$, if one notes that the variable $t$ in the expression 
$\alpha_s\left(\frac{xP}{zt}\right)$ was related to the virtuality 
$Q^2$ of the parent parton and should therefore be replaced
by the age variable $\tau=\frac{xP}{z Q^2}$. This immediately yields:
\begin{eqnarray}
\left(\frac{\partial \,g}{\partial t}\right)_{\rm branch} &=&
-\;\frac{1}{2}\;\int_0^1 d z 
\;\frac{\overline{\alpha_s}(\tau)}{2 \pi \tau} 
\;\gamma_{g \rightarrow gg}(z)\;
g(x,\tau,t)
\nonumber \\
& & \;\;+\;
\int_0^1 \frac{dz}{z}
\;\frac{\alpha_s\left(\frac{xP}{z \tau}\right)}{2 \pi \tau} 
\;\theta \left(\frac{xP}{z \tau} - \mu_0^2\right)
\;\gamma_{g \rightarrow gg}(z)\;
g\left(\frac{x}{z},\tau,t\right)
\;\;.
\end{eqnarray}

(iii)
The scattering term is somewhat more complicated. We begin by noting 
that the momentum fractions of two partons before $(x_1,x_2)$ and after 
$(x_1^\prime, x_2^\prime)$ scattering are related
 by
\begin{equation}
x_{1;2}^\prime\;\,=\;\, \frac{x_1 + x_2}{2} \;\pm\; 
\sqrt{\frac{(x_1 - x_2)^2}{4} \;-\; \frac{p_\perp^2}{P^2}}
\;\;,
\end{equation}
or
\begin{equation}
x_1\;=\; x_1^\prime \;+\;\frac{p_\perp^2}{(x_1^\prime - x_2) P^2}
\;\;\;,
\;\;\;\;\;\;\;\;\;
x_2\;=\; x_2^\prime \;+\;\frac{p_\perp^2}{(x_2^\prime - x_1) P^2}
\;\;.
\label{x12}
\end{equation}
Since only gluons with opposite directions of propagation are allowed 
to scatter in the parton picture (c.f. discussion in Sec. 2.1),
the total scattering rate of cascading gluons is given by
\begin{equation}
w \;=\;  \int_{0}^1 dx_1 \int_0^\infty d \tau_1\int_{-1}^0 dx_2 
\int dp_\perp^2 \; g(x_1,\tau_1,t) \;\hat g(x_2)
\rho_N\;{d\hat\sigma_{gg\rightarrow gg}\over dp_\perp^2} 
\;\;,
\end{equation}
where $\rho_N$ is the nucleon density in the medium.  The loss term 
due to the scattering of a cascading gluon is therefore
\begin{equation}
\left(\frac{\partial \,g}{\partial t}\right)_{\rm scatt}^{\rm (loss)} \;=\;
-\;\int_{-1}^0 dx_2 \int d p_\perp^2 \;g(x,\tau,t) \;\hat g(x_2) \;
{d\hat\sigma_{gg\rightarrow gg}\over dp_\perp^2} 
\;\rho_N
\;\;.
\end{equation}
In the gain term we have to distinguish between ``soft'' and ``hard'' collisions, 
as compared with the virtuality of the incoming cascading gluon 
$Q_1^2 = x_1P/\tau_1$. If $Q_1^2 > p_\perp^2$, the scattering parton will 
keep its virtuality, because the collision cannot be resolved from the 
previous interaction that originally kicked the parton off mass shell. If 
$Q_1^2 < p_\perp^2$, the scattered parton will acquire maximum virtuality 
$p_\perp^2$, corresponding to an age $\tau = \vert x\vert P/p_\perp^2$.  
The gluon scattered out of the nuclear medium, however,  always acquires 
maximum virtuality $p_\perp^2$, because it was space-like before the 
interaction. Therefore: 
\begin{eqnarray}
\left(\frac{\partial \,g}{\partial t}\right)_{scatt}^{(gain)} &=&
\int_0^1 dx_1 \int _0^\infty d \tau_1 \int_{-1}^0 dx_2 \int d p_\perp^2 
\;g(x_1,\tau_1,t) \;\hat g(x_2) \;
{d\hat\sigma_{gg\rightarrow gg}\over dp_\perp^2} 
\;\rho_N
\;\delta(x_1^\prime - x)
\nonumber \\
& & \;\;\;\;
\times
\;\left[ \delta\left(\tau -\frac{|x|P}{p_\perp^2}\right)
\;\left(1 \,+\,\theta\left(p_\perp^2 - \frac{x_1P}{\tau_1}\right)\right)
\;+\;\delta (\tau - \tau_1)\,\theta\left(\frac{x_1P}{\tau_1} - p_\perp^2\right)
\right]
\;\;.
\end{eqnarray}
For the first term in the brackets we can perform the $p_\perp^2$-integration,
for the second term, the $\tau_1$-integration;
the $x_1$-integration collapses, yielding
\begin{equation}
x_1\;=\; x \;+\;\frac{p_\perp^2}{(x - x_2) P^2}
\end{equation}
according to eq. (\ref{x12}). This leaves us with
\begin{eqnarray}
\left(\frac{\partial \,g}{\partial t}\right)_{\rm scatt}^{\rm (gain)} &=&
\int_{-1}^0 dx_2 \;\hat g(x_2) \;\frac{xP}{\tau^2}\;
\left. {d\hat\sigma_{gg\rightarrow gg}\over dp_{\perp}^2}  
\right|_{p_\perp^2 = \frac{|x|P}{\tau}} \;\rho_N
\;\int_0^\infty d \tau_1 \; g(x_1,\tau_1,t)
\;\left[1+\theta\left(\frac{\tau_1}{\tau} - \frac{x_1}{\vert x\vert} 
\right)\right]
\nonumber \\
& & \;\;+\;
\int_{-1}^0 dx_2 \int d p_\perp^2\; g(x_1,\tau, t) \;\hat g(x_2)\; 
\frac{d\hat\sigma_{gg\rightarrow gg}}{dp_\perp^2} \;\rho_N
\;\theta\left(\frac{x_1P}{\tau} - p_\perp^2\right)
\;\;.
\end{eqnarray}
Note that since $x_1$ itself depends on $p_\perp^2$ (for fixed $x$ and 
$x_2$), the step function in the last term really imposes the integration 
limit
\begin{equation}
\theta\left(\frac{x_1P}{\tau} - p_\perp^2\right)
\;=\;
\theta\left(\frac{x (x-x_2) P^2}{(x-x_2) P \tau - 1 } \,-\,p_\perp^2\right)
\;\;.
\end{equation}
(Of course, the integration range is also limited by the condition $x_1 > 0$.)

(iv)
Finally we turn to the fusion processes, where the invariant mass $M^2$ of the
produced off-shell parton replaces the momentum transfer
$p_\perp^2$ as virtuality scale.
Since only those gluons that have reversed their direction of propagation, 
i.e. with $x < 0$, can fuse with medium gluons (c.f. discussion in Sec. 2.1), 
the total gluon fusion rate is given by
\begin{equation}
\overline{w} \;=\; \int_{-1}^0 dx_1 \int_0^\infty d \tau_1 \int_{-1}^0 
dx_2 g(x_1,\tau_1,t) \; \hat g(x_2) \;\rho_N
\; \Gamma_{gg\to g}(x_1,x_2,x_1+x_2) \;{4\pi\alpha_s(M^2)\over M^2} ,
\end{equation}
where the fusion probability $\Gamma_{a b\rightarrow c}$ is related to 
the branching functions
$\gamma_{a \rightarrow bc}$ by \cite{close}
\begin{equation}
\Gamma_{a b \rightarrow c}(x_1, x_2, x_3) \;=\; c_{a b \rightarrow c} \;
\frac{ x_1 x_2}{x_3^2} \; \gamma_{a \rightarrow b c}\left(\frac{x_1}{x_3}\right)
\;=\; c_{a b \rightarrow c} \;
\frac{ x_1 x_2}{x_3^2} \; \gamma_{a \rightarrow c b}\left(\frac{x_2}{x_3}\right)
\;\;,
\label{Gamma}
\end{equation}
where $x_3=x_1+x_2$, and the factors in front of $\gamma_{a \rightarrow bc}$ 
arise from the difference of phase-space and flux factors for fusions 
compared to branchings.  The color factors $c_{a b\rightarrow c}$ are 
$c_{g g \rightarrow g} = 1/8$,  $c_{q g \rightarrow q} = 1/8$,
$c_{q \bar q \rightarrow g} = 8/9$.  Hence, we have
\begin{equation}
\Gamma_{gg\to g}(x_1,x_2,x_1+x_2) \;=\; {x_1x_2\over 8(x_1+x_2)^2}
\gamma_{g\to gg}\left({x_1\over x_1+x_2}\right) ,
\end{equation}
and the invariant mass, neglecting the virtuality of the medium parton [see eq. (\ref{m2})], is
\begin{equation}
M^2 \;=\; \frac{|x_1|\,P}{\tau_1}\, \left(1 \,+\, \frac{x_2}{x_1}\right)
\;=\; \frac{|x_1+x_2|\ \,P}{\tau_1}
\;\;.
\label{m22}
\end{equation}
Here $x_1 < 0$ and $\tau_1$ denote the momentum fraction and 
virtuality of the shower parton fusing with a parton from the nuclear 
medium, which has momentum fraction $x_2<0$.  Since the virtuality of 
the cascading gluon before fusion was $Q_1^2 = |x_1| P/\tau_1$, we see 
that the  {\it same} age $\tau_1$ describes its virtuality correctly 
before {\it and} after fusion.  In other words, the variable $\tau$ 
remains unchanged by fusion processes.  The loss and gain term from 
fusion processes are therefore:
\begin{eqnarray}
\left(\frac{\partial g}{\partial t}\right)_{fus} &=&
-\;\int_{-1}^0 dx_2 \;g(x,\tau,t)\;\hat g(x_2) \;\rho_N\;
\Gamma_{gg\to g}(x,x_2,x+x_2)
\left. \;{4\pi\alpha_s(M^2)\over M^2} \right|_{M^2 = |x+x_2| P/\tau}
\nonumber \\
& & \;\;+
\;\int_{-1}^0 dx_2\; g(x-x_2,\tau,t) \;\hat g(x_2) \;\rho_N\;
\Gamma_{gg\to g}(x-x_2,x_2,x)
\left. \;{4\pi\alpha_s(M^2)\over M^2} \right|_{M^2 = |x| P/\tau}
\;\;.
\end{eqnarray} 

We finally combine all our results to write down the full evolution 
equation for the off-shell gluon distribution:
\begin{eqnarray}
{\partial\over\partial t}g(x,\tau,t) \;=\;
& - &{\partial\over\partial t}g(x,\tau,t)
\nonumber \\
& - &
{1\over 2} \;\int_0^1 dz \;\gamma_{g\rightarrow gg}(z) \;\frac{\overline{\alpha_s}(\tau)}{2 \pi \tau}
\; g(x, \tau, t)
\nonumber \\
& + &
\;\int_0^1 \frac{dz}{z} \frac{\alpha_s\left(\frac{x P}{z \tau}\right)}{2 \pi \tau}
\;\theta \left(\frac{x P}{z \tau} - \mu_0^2 \right) \;\gamma_{g\rightarrow gg}(z)
\; g\left(\frac{x}{z}, \tau, t\right)
\nonumber \\
& - &
\;\int_{-1}^0 dx_2 \;\hat g(x_2) \;\int d p_\perp^2 \;
{d\hat\sigma_{gg\rightarrow gg}\over dp_\perp^2} 
\;\rho_N
\;g(x,\tau,t)
\nonumber \\
& + &
\frac{xP}{\tau^2}\;
\int_{-1}^0 dx_2 \;\hat g(x_2) \;
\left. {d\hat\sigma_{gg\rightarrow gg}\over dp_\perp^2}  
\right|_{p_\perp^2 = \frac{|x|P}{\tau}} 
\nonumber \\
& & 
\;\;\;\;\;\;\;\;\;
\times 
\;\rho_N
\int_0^\infty d \tau_1 \; g(x_1,\tau_1,t)
\;\left[1 \,+\,\theta\left({\tau_1\over\tau} - \frac{x_1}{\vert x\vert}
\right) \right]
\nonumber \\
& + & 
\;\int_{-1}^0 dx_2 \;\hat g(x_2)\;\int^{x_1 P/\tau} d p_\perp^2\; 
{d\hat\sigma_{gg\rightarrow gg}\over dp_\perp^2}
\;\rho_N
\;g(x_1,\tau,t)
\nonumber \\
& - & 
\;\int dx_2 \;\hat g(x_2) \;g(x,\tau,t)\;\rho_N\;
\Gamma_{gg\to g}(x,x_2,x+x_2)
\left. \;{4\pi\alpha_s(M^2)\over M^2} \right|_{M^2 = |x+x_2| P/\tau}
\nonumber \\
& + &
\;\int_{-1}^0 dx_2\; \hat g(x_2)\;g(x-x_2,\tau,t) \;\rho_N\;
\Gamma_{gg\to g}(x-x_2,x_2,x)
\left. \;{4\pi\alpha_s(M^2)\over M^2} \right|_{M^2 = |x| P/\tau}
\;\;.
\label{geveq}
\end{eqnarray} 
Note that the modified equation no longer is a differential equation in
the variable $(\ln t)$, but rather in $t$ directly, because the presence
of interactions with the medium defines a characteristic timescale
$(\sigma\rho)^{-1}$, the mean free time between scatterings, which
breaks the scale invariance of the fragmentation cascade.
\medskip

\noindent {\bf 2.5 Transverse momentum spread}
\smallskip

In order to investigate the development of lateral spread perpendicular to
the parton shower axis, one has to study the transverse momentum dependence 
of the gluon distribution. Recall that the distributrion $g(x,\tau,t)$ is the
zeroth moment of the full gluon phase-space distribution  $g(x,k_\perp^2, 
\tau,t)$ [c.f. eq. (\ref{g0m})],
\begin{equation}
g(x,\tau, t) \;\equiv\; \int dk_\perp^2 \; g(x,k_\perp^2,\tau,t)
\;\;.
\label{gzm}
\end{equation}
Instead of keeping the transverse momentum $k_\perp$ as an independent
variable in the gluon distribution function we here will only follow
its average growth due to branching, scattering and fusion processes. In order
to do this, we introduce the mean squared transverse momentum distribution 
$\pi_g(x,\tau,t)$ of gluons as the first moment in $k_\perp^2$, i.e.
\begin{equation}
\pi_g(x,\tau,t) \;\equiv\; \int dk_\perp^2 \;k_\perp^2 \;g(x,k_\perp^2,\tau,t) 
\;\;.
\end{equation}
The evolution equation for $\pi_g(x,\tau, t)$ is easily derived in analogy to
the equation (\ref{geveq}) for the gluon distribution function $g(x,\tau,t)$.
Each loss and gain term is to be weighted with the transverse momentum squared
$k_\perp^2$ as it changes by accumulating a certain $p_\perp^2$ generated 
in the branching, scattering, or fusion processes.  Since the loss terms in 
(\ref{geveq}) generally also represent the loss of partons out of a range 
between $k_\perp^2$ and $k_\perp^2 + d k_\perp^2$, one has simply to replace 
$g(x,\tau,t)$ by $\pi_g(x,\tau,t)$ in these terms.  The gain terms on the 
other hand receive different contributions associated with the different 
kinematics of branching, scattering and fusion.

(i) For a branching process $k_{j-1} \rightarrow k_j + k_j^\prime$,
using (\ref{ptcons},\ref{econs}) and
assuming $Q_j^2,Q_j^{\prime 2} \ll Q_{j-1}^2$, one finds:
\begin{eqnarray}
k_{\perp j}^2 & \approx & z_j^2 \, k_{\perp j-1}^2 \;+\; z_j ( 1 - z_j) \; 
Q_{j-1}^2
\nonumber \\
k_{\perp j}^{\prime\,2} &\approx & (1-z_j)^2 \, k_{\perp j-1}^2 \;+\; z_j 
( 1 - z_j) \; Q_{j-1}^2
\;\;.
\end{eqnarray}
(ii) For a scattering process between a shower parton with $k_{j-1}$ and 
a nuclear parton, involving a squared transverse momentum exhange of 
$p_{\perp j}^2$, with the two corresponding outgoing partons carrying
$k_j$ and  $k_j^\prime$, respectively, we have
\begin{equation}
\vec k_{\perp j}\;=\; \vec k_{\perp j-1}\;+\; \vec p_{\perp j}
\;\;\;,\;\;\;\;\;\;\;\;\;\;
\vec k_{\perp j}^\prime\;=\; -\; \vec p_{\perp j}
\end{equation}
since the nuclear parton initially has no transverse momentum, and
\begin{equation}
p_{\perp j}^2\;=\;  k_{\perp j}^2 \;-\; k_{\perp j-1}^2 \;=\; k_{\perp j}^{\prime\, 2}
\;\;.
\end{equation}

(iii) For a fusion process between a cascade parton with $k_{j-1}$ and a 
parton from the nuclear medium yielding the compound state with $k_j$, the 
condition is $\vec k_{\perp j}= \vec k_{\perp j-1}$, because as in the 
scattering case, the nuclear parton initially carries only longitudinal 
but no transverse momentum, hence
\begin{equation}
k_{\perp j}^2\;=\;  k_{\perp j-1}^2
\;\;.
\end{equation}

Using these kinematic conditions on the change of transverse momentum in 
the interactions, we find the evolution equation for the
first moment in $k_\perp^2$ of the off-shell gluon distribution:
\begin{eqnarray}
{\partial\over\partial t}\pi_g(x,\tau,t) \;=\;
& - &{\partial\over\partial t}\pi_g(x,\tau,t)
\nonumber \\
& - &
{1\over 2} \;\int_0^1 dz \;\gamma_{g\rightarrow gg}(z) \;
\frac{\overline{\alpha_s}(\tau)}{2 \pi \tau}
\; \pi_g(x, \tau, t)
\nonumber \\
& + &
\;\int_0^1 \frac{dz}{z} \frac{\alpha_s\left(\frac{x P}{z \tau}\right)}
{2 \pi \tau} \;\theta \left(\frac{x P}{z \tau} - \mu_0^2 \right) \;
\gamma_{g\rightarrow gg}(z)
\nonumber \\
& & 
\;\;\;\;\;\;\;\;\;
\times \;
\left[ z (1-z)\,\frac{xP}{z\tau}
\; g\left(\frac{x}{z}, \tau, t\right)
\; +\; z^2\, \pi_g\left(\frac{x}{z}, \tau, t\right)
\right]
\nonumber \\
& - &
\;\int_{-1}^0 dx_2 \;\hat g(x_2) \;\int d p_\perp^2 \;
{d\hat\sigma_{gg\rightarrow gg}\over dp_\perp^2} 
\;\rho_N
\;\pi_g(x,\tau,t)
\nonumber \\
& + &
\frac{xP}{\tau^2}\;
\int_{-1}^0 dx_2 \;\hat g(x_2) \;
\left. {d\hat\sigma_{gg\rightarrow gg}\over dp_\perp^2}  \right|_{p_\perp^2 = \frac{|x|P}{\tau}} 
\nonumber \\
& & 
\;\;\;\;\;\;\;\;\;
\times 
\;\rho_N
\int_0^\infty d \tau_1 \; 
\left\{
p_\perp^2 \, g(x_1,\tau_1,t)
\;\left[1 \,+\,\theta\left(\frac{\tau_1}{\tau} - \frac{x_1}{\vert
x\vert} \right) \right]
\right.
\nonumber \\
& & \;\;\;\;\;\;\;\;\;\;\;\;\;
\left.
\;+\;
\pi_g(x_1,\tau_1,t) \;\theta\left(\frac{\tau_1}{\tau} - 
\frac{x_1}{\vert x\vert} \right)
\;\right\}
\nonumber \\
& + & 
\;\int_{-1}^0 dx_2 \;\hat g(x_2)\;\int^{x_1P/\tau} d p_\perp^2\; 
{d\hat\sigma_{gg\rightarrow gg}\over dp_\perp^2}
\;\rho_N
\;\left[ p_\perp^2\;g(x_1,\tau,t)\;+\;\pi_g(x_1,\tau,t) \right]
\nonumber \\
& - & 
\;\int dx_2 \;\hat g(x_2) \;\pi_g(x,\tau,t)\;\rho_N\;
\Gamma_{gg\to g}(x,x_2,x+x_2)
\left. \;{4\pi\alpha_s(M^2)\over M^2} \right|_{M^2 = |x+x_2| P/\tau}
\nonumber \\
& + &
\;\int_{-1}^0 dx_2\; \hat g(x_2)\;\pi_g(x-x_2,\tau,t) \;\rho_N\;
\Gamma_{gg\to g}(x-x_2,x_2,x)
\left. \;{4\pi\alpha_s(M^2)\over M^2} \right|_{M^2 = |x| P/\tau}
\;\;.
\label{peveq}
\end{eqnarray} 
\medskip

\noindent {\bf 3. THE COUPLED EVOLUTION EQUATIONS FOR QUARKS,
ANTIQUARKS  AND GLUONS}
\medskip

In this section we will extend the previous derivation of the gluon
evolution
to the coupled system of gluons $g$, quarks $q_i$ and antiquarks
$\bar q_i$ with flavors $i=1, \ldots f$. The only essential difference to the
purely gluonic case is that now the parton cascade evolves via
a number of different branching, fusion and scattering subprocesses
which couple the gluons with the quarks and antiquarks.
Let us denote the rates for the various interactions of the cascading
partons 
due to $1\rightarrow 2$ branchings, $2 \rightarrow 1$ fusions, and 
$2 \rightarrow 2$ scatterings by
($a = q_i, \bar q_i, g$)
\begin{equation}
R_a^{(m \rightarrow m^\prime)}(x,\tau,t)\;\equiv \; 
\left(\frac{\partial}{\partial t}\,
\,x\,a(x,\tau,t) \right)_{{\rm processes} \,m\rightarrow m^\prime}
\;\;,
\end{equation}
i.e. as the change of the $x$-weighted parton densities, integrated 
over $k_\perp^2$, 
\begin{equation}
x\;a(x, \tau,t) \;=\; 
\int dk_\perp^2 \;x\,a(x, k_\perp^2, \tau, t) 
\;\;.
\end{equation}
To evaluate the interaction rates, 
$R_a^{(m \rightarrow m^\prime)}(x,\tau,t)$,
we use the well known lowest order perturbative QCD expressions for the 
branching amplitudes \cite{ap,collins},
fusion amplitudes \cite{close,mulqiu},
and the parton-parton cross-sections
\cite{fields,combridge,svetitsky}, respectively, and implement those
in the formalism described in the previous Section.
The corresponding Feynman diagrams are depicted in Figs. 7-9.

For the following we introduce the {\it parton momentum densities}
$Q_i$ ($\bar Q_i$) and $G$
\begin{equation}
Q_i(x, \tau,t) \;=\; 
\int dk_\perp^2 \;x\,q_i(x, k_\perp^2, \tau,t) \;\;\;,\;\;\;\;\;\;\;\;\;\;\;
G(x, \tau, t) \;=\;
\int dk_\perp^2 
x \,g(x, k_\perp^2, \tau, t)
\;\;.
\label{QG}
\end{equation}
i.e. the parton number densities $q_i$ ($\bar q_i$) and $g$ weighted with 
the longitudinal momentum fraction $x$, where $i = 1, \ldots f$ denotes the
quark flavors. Furthermore, for brevity we will 
employ the notation:
\begin{eqnarray}
\alpha_s \left(\frac{xP}{\tau}\right)&=&
\frac{12 \pi}{(33-2 f) \ln (\frac{|x| P}{\tau\Lambda^2})}
\nonumber \\
\overline{\alpha_s}\left(\tau\right) &\equiv&
\frac{
\;\int_0^1 d x \;
\alpha_s\left(\frac{\vert x\vert P}{\tau}\right)\;
\theta\left(\frac{\vert x\vert P}{\tau} \,-\,\mu_0^2\right)
\; \left(g\left(x,\tau,t\right)\,+\,\sum_j [q_j(x,\tau,t) + 
\bar q_j(x,\tau,t)] \right)
}{
\int_0^1 d x \,\left( g(x,\tau,t) \,+\, \sum_j [q_j(x,\tau,t) + \bar q_j(x,\tau,t)] \right)
}
\nonumber \\
\overline{\kappa}(\tau) &\equiv&
\frac{\overline{\alpha_s}(\tau)}{2 \pi \tau}
\nonumber\\
\kappa(x,\tau) &\equiv&
{\alpha_s(\vert x\vert P/\tau)\over 2\pi \tau} 
\theta\left({\vert x\vert P\over \tau}-\mu_0^2\right) 
\nonumber\\
\zeta(x,\tau) &\equiv&
\left. {4\pi\alpha_s(M^2)\over M^2} \right|_{M^2 = \vert x\vert P/\tau} 
\;\;.
\end{eqnarray}
Finally we recall that $\tau = \vert x\vert P/Q^2$
denotes the life-time of a parton with longitudinal momentum fraction 
$x$ and virtuality $Q^2$, the variable $z=x/x^\prime$ is the fraction of 
$x$-values of daughter to mother partons in branchings and fusions,
$M^2$ is the invariant mass squared of two fusing partons, and
$p_\perp^2$ refers to the relative transverse momentum squared exchanged 
in parton-parton scatterings.
\medskip

\noindent
{\bf 3.1 Branching processes}

The net change
of the quark number densities 
due to the branching processes shown in Fig. 7 a)
is obtained by adding up the gain term 
due to quarks of momentum fraction
$x_1 > x$ having radiated a gluon of momentum fraction
$x_1 - x$, the loss term, identifying all those quarks that
had momentum fractions $x$ before radiating a gluon of momentum fraction
less than $x$, and 
the additional gain term that arises from gluons with momentum fraction 
$x_1 > x$ decaying in a $q_i \bar q_i$ pair with momentum fractions $x$ 
and $x_1-x$, respectively.  Correspondingly, the net change of the 
antiquark number distributions due to branchings is given by replacing 
$q_i$ by $\bar q_i$.

The result for the branching rates of quarks (and analogous for antiquarks) is: 
\begin{equation}
R_{q_i}^{(1 \rightarrow 2)} (x,\tau, t) \;=\;
- \;\hat A \,Q_i \; + \; \hat B \, G
\;\;,
\end{equation}
where
\begin{eqnarray}
- \;\hat A \,Q_i &=&
-\; 
\int_0^1 d z \left[
Q_i(x,\tau,t) \,\overline{\kappa} (\tau)\,-\,Q_i\left(\frac{x}{z}, 
\tau,t\right)\,
\kappa\left(\frac{x}{z},\tau\right) \right]
\; \gamma_{q \rightarrow q g}(z) 
\nonumber
\\
 \;\hat B \,G &=&
\int_0^1 d z 
\; G\left(\frac{x}{z}, \tau,t\right) 
\;\kappa\left(\frac{x}{z},\tau\right)
\; \gamma_{g \rightarrow q \bar q}(z)
\;\;.
\label{bq}
\end{eqnarray}
The branching functions $\gamma_{a\rightarrow bc}(z)$ are
given by (\ref{gamma}).

The change of the gluon distributions is similarly obtained by adding 
the contributions of Fig. 7 b), namely the gain of gluons with momentum 
fraction $x$ due to gluon emission by gluons with momentum fraction 
$x_1 > x$, the loss term due to gluon emission by gluons with momentum 
fraction $x$, the loss term due to gluon decay into $q_i \bar q_i$ is, 
summed over all quark flavors $i$, and, the gain term due to radiation 
of gluons with momentum fraction $x$ by quarks and antiquarks with 
momentum fractions $x_1 > x$.  We then obtain for the branching rates 
of gluons: 
\begin{equation}
R_{g}^{(1 \rightarrow 2)} (x,\tau, t) \;=\;
- \;\hat C \,G \; - \; \hat D \, G
+ \;\sum_{j=1}^f \left(\frac{}{} \hat E \,Q_j \; + \; \hat E \, \bar Q_j \right)
\;\;,
\end{equation}
where
\begin{eqnarray}
- \;\hat C \, G &=&
-\; 
\int_0^1 d z \left[
\frac{1}{2}\,
G(x,\tau,t) \,\overline{\kappa}(\tau)\,-\,G\left(\frac{x}{z}, 
\tau,t\right)\,\kappa (x,\tau) \right]
\; \gamma_{g \rightarrow g g}(z) 
\nonumber
\\
- \;\hat D \, G &=&
-
\;
f \;
G(x,\tau,t) 
\;
\int_0^1 d z 
\;\overline{\kappa}(\tau)
\; \gamma_{g \rightarrow q \bar q}(z)
\nonumber
\\
\;\hat E \,Q_j &=&
\int_0^1 d z 
\; Q_j\left(\frac{x}{z},\tau,t\right) 
\;\kappa (x,\tau)
\; \gamma_{q \rightarrow q  g}(z)
\nonumber
\\
\;\hat E \,\bar Q_j &=&
\int_0^1 d z 
\;\bar Q_j\left(\frac{x}{z},\tau,t\right) 
\;\kappa (x,\tau)
\; \gamma_{q \rightarrow q  g}(z)
\;\;.
\label{bg}
\end{eqnarray}

\noindent
{\bf 3.2 Fusion processes}
\smallskip

The parton-parton fusion processes manifestly alter the parton number 
densites in a similar manner as the branching processes, provided the 
quark and gluon densities of the nuclear medium (labeled with a ``hat'') 
are sufficiently dense and the probability for a parton to absorb one 
of the nuclear partons becomes significant.  Recall that we do not 
consider here fusions among the shower partons or among the nuclear 
partons themselves. To obtain the fusion rates, we use the expressions 
for the fusion probabilities $\Gamma_{ab\rightarrow c}$ given in (\ref{Gamma})
in terms of the branching functions $\gamma_{a\rightarrow bc}$.

The net change of the quark number distributions 
due to fusions with partons in
the nuclear medium (labeled by a ``hat'') 
is balanced by the gain and loss of quarks with momentum fraction $x$ due to
the processes shown in Fig. 8 a).
These are the gain of quarks through fusions of a quark with $x_1 < x$ and 
a gluon with $x_2 < x$  such that $x_1+x_2 = x$, the loss of quarks with 
longitudinal momentum fraction $x$ due to fusions with gluons of $x_2$,
and, the  loss of quarks with fraction $x$ due to $q_i 
\bar q_i$-annihilation into gluons.  The corresponding change in the 
antiquark number distributions is obtained by interchanging $q_i$ and 
$\bar q_i$.

The result for the fusion rates of quarks or antiquarks is:
\begin{equation}
R_{q_i}^{(2 \rightarrow 1)} (x, \tau, t) \;=\;
- \;\hat A^\prime [\hat G] \,Q_i \; - \; \hat B^\prime [\hat{\bar{Q}}_i] \,Q_i
\;\;,
\end{equation}
where, in contrast to $\hat A$ and $\hat B$ in (\ref{bq}), 
the integral operators $\hat A^\prime$ and $\hat B^\prime$
are functionals of the densities of quarks $\hat Q$ ($\hat{\bar Q}$) and 
gluons $\hat G$ of the nuclear medium
in which the cascading particles evolve. 
The integral operators $A^\prime$ and $B^\prime$ are obtained as:
\begin{eqnarray}
- \;\hat A^\prime [\hat G] \,Q_i &=&
-\; 
\frac{1}{8}\;
\int_0^1 d z \left[
Q_i(x,\tau, t) \hat G\left(\frac{x (1-z)}{z}\right)
\,\zeta\left(\frac{x}{z}, \tau \right)
\right.
\nonumber
\\
& & \;\;\;\;\;\;\;\;
\left.
\frac{}{}
\,-\,\frac{1}{x^2}\,Q_i\left(x z, \tau, t \right)\, \hat G\left( x (1-z)\right) \right]
\; \gamma_{q \rightarrow q g}(z) 
\,\zeta\left(x, \tau \right)
\nonumber
\\
- \;\hat B^\prime [\hat{\bar{Q}}_i] \,Q_i &=&
-\,\frac{8}{9}\;
\int_0^1 d z 
\;Q_i(x,\tau,t) \hat{\bar{Q}}_i\left(\frac{x (1-z)}{z} \right)
\; \gamma_{g \rightarrow q \bar q}(z)
\,\zeta\left(\frac{x}{z}, \tau \right)
\;\;.
\end{eqnarray}

The gluon number distributions 
receive modifications from the fusion processes depicted in Fig. 8 b).
There is the  gain of gluons with momentum fraction $x$ from fusions of two gluons
with  $x_1 < x$ and $x_2 < x$ 
such that $x_1+x_2 = x$,
the loss of gluons with $x$ due to fusions with other gluons from the medium,
the  gain of gluons with fraction $x$ due to $q_i \bar q_i$-annihilation,
and, the loss of gluons with fraction $x$ due to absorption by quarks
or antiquarks.
We obtain the following  result for the fusion rates of gluons: 
\begin{eqnarray}
R_{g}^{(2 \rightarrow 1)} (x, \tau, t) &=&
- \;\hat C^\prime [\hat G] \,G \; - \; \sum_{j=1}^f \left(\frac{}{}\hat D^\prime [\hat Q_j] \, G
+ \; \hat D^\prime [\hat{\bar{Q}}_j] \, G \right)
\nonumber
\\
& & \;\;\;\;
+ \; \sum_{j=1}^f\left(\frac{}{}\hat E^\prime [\hat{\bar{Q}}_j] \,  Q_j
+ \; \hat E^\prime [\hat  Q_j] \,\bar Q_j\right)
\;\;,
\end{eqnarray}
where 
\begin{eqnarray}
- \;\hat C^\prime [\hat G] \, G &=&
-\; 
\frac{1}{8}\;
\int_0^1 d z \bigg[
\,G(x,\tau,t) \hat G \left( \frac{x (1-z)}{z} \right)
\,\zeta\left(\frac{x}{z}, \tau \right) \nonumber \\
&&\qquad \qquad -\frac{1}{x^2}\,G\left(x z, \tau,t\right)\, 
\hat G\left( x (1-z) \right) 
\,\zeta\left(x, \tau \right)
\bigg] \
\; \gamma_{g \rightarrow g g}(z) 
\nonumber
\\
- \;\hat D^\prime [\hat Q_j] \,G &=&
-\;
\frac{1}{8}\,
G(x,\tau,t)\;
\int_0^1 d z 
\; \hat{Q}_j\left(\frac{x (1-z)}{z} \right)
\; \gamma_{q \rightarrow q g}(z)
\,\zeta\left(\frac{x}{z}, \tau \right)
\nonumber \\
- \;\hat D^\prime [\hat{\bar Q}_j] \,G &=&
-\;
\frac{1}{8}\,
G(x,\tau,t)\;
\int_0^1 d z 
\; \hat{\bar{Q}}_j\left(\frac{x (1-z)}{z} \right)
\; \gamma_{q \rightarrow q g}(z)
\,\zeta\left(\frac{x}{z}, \tau \right)
\nonumber
\\
\;\hat E^\prime [\hat{\bar{Q}}_j] \, Q_j &=&
\frac{4}{9}\,
\frac{1}{x^2}\;
\int_0^1 d z 
\; Q_j\left(x (1-z)\right) \hat{\bar{Q}}_j \left(x z \right)
\; \gamma_{g \rightarrow q \bar q}(z)
\,\zeta\left(x, \tau \right)
\nonumber
\\
 \;\hat E^\prime [\hat Q_j] \, \bar Q_j &=&
\frac{4}{9}\,
\frac{1}{x^2}\;
\int_0^1 d z 
\;\bar{Q}_j\left(x (1-z)\right) \hat{Q}_j\left(x z \right)
\; \gamma_{g \rightarrow q \bar q}(z)
\,\zeta\left(x, \tau \right)
\;\;.
\end{eqnarray}

\noindent
{\bf 3.3 Scattering processes}
\smallskip

Finally, the collision rates for elastic scatterings of the cascading partons with
the partons in the nuclear backgound medium with nuclear density $\rho_N$ receive
various contributions which are diagramatically shown in Fig. 9. 
Again, we only account for interactions of the parton cascade with the medium, i.e. those
parton collisions that involve a shower parton and a nuclear parton, the latter of which after the scattering
becomes a time-like excitation and is added to the cascade.
To compress the expressions for the scattering rates, we introduce the following
function:
\begin{eqnarray}
\hat S_{a b} [\hat A] \, B  &\equiv&
-\;\int_{-1}^0 dx_2 \;\hat A(x_2) \;\int d p_\perp^2 \;
\frac{1}{x x_2}\,{d\hat\sigma_{ab\rightarrow ab}\over dp_\perp^2} 
\;\rho_N
\;B(x,\tau,t)
\nonumber \\
& + &
\frac{xP}{\tau^2}\;
\int_{-1}^0 dx_2 \;\hat A(x_2) \;
\frac{1}{x_1 x_2}\,
\left. {d\hat\sigma_{ab\rightarrow ab}\over dp_\perp^2}  \right|_{p_\perp^2 = \frac{|x|P}{\tau}} 
\nonumber \\
& & \;\;\;\;\;\;\;
\times \;
\;\rho_N
\int_0^\infty d \tau_1 \; B(x_1,\tau_1,t)
\;\left[1 \,+\theta\left(\frac{\tau_1}{\tau} - \frac{x_1}{\vert x\vert}
\right) \right]
\nonumber \\
& + & 
\;\int_{-1}^0 dx_2 \;\hat A(x_2)\;\int^{x_1 P/\tau} d p_\perp^2\; 
\frac{1}{x_1 x_2}\,
{d\hat\sigma_{ab\rightarrow ab}\over dp_\perp^2}
\;\rho_N
\;B(x_1,\tau,t)
\;\;,
\label{Sab}
\end{eqnarray}
where $A, B = G, Q_j, \bar Q_j$ and the ``hat'' labels as before the 
nuclear parton distributions, whereas the distributions without ``hat'' 
refer to the cascading partons.

For the quarks (and similar antiquarks) we have the processes
$q_i g\rightarrow q_i g$, $q_i q_j \rightarrow q_i q_j$ and $q_i 
\bar q_j \rightarrow q_i \bar q_j$.  Hence,
\begin{equation}
R_{q_i}^{(2 \rightarrow 2)} (x, \tau,t) \;=\;
\hat S_{q_i g} [\hat Q_i] \,G 
\;+ \; \sum_{j=1}^f\left(\frac{}{} \hat S_{q_i q_j} [\hat Q_i] \, Q_j
\; + \; \hat S_{q_i \bar q_j} [\hat Q_i] \, \bar Q_j \right)
\end{equation}

On the other hand, for gluons the contributing processes are
$g g\rightarrow g g$, $g q_j \rightarrow g q_j$ and $g \bar q_j 
\rightarrow g \bar q_j$.  Consequently,
\begin{equation}
R_{g}^{(2 \rightarrow 2)} (x, \tau, t) \;=\;
\hat S_{g g} [\hat G] \,G
\;+ \; \sum_{j=1}^f \left(\frac{}{}\hat S_{g q_j} [\hat G] \, Q_j
\; + \; \hat S_{g \bar q_j} [\hat G] \, \bar{Q}_j \right)
\end{equation}

The parton-parton cross-sections that enter the expressions $\hat S_{ab} 
[\hat A]\,B$, eq. (\ref{Sab}), i.e., $d \hat \sigma_{a b\rightarrow c d}/d 
q_\perp^2$, for massless partons are related to the squared scattering 
amplitudes $\vert \overline{\cal M}_{a b \rightarrow c d}\vert ^2$, averaged
over initial spin and color states and summed over the final states, 
by
\begin{equation}
\frac{d \hat \sigma_{a b \rightarrow c d}(\hat s, p_\perp^2)}{d p_\perp^2}
\;=\;
{\cal D}_{ab}\,{\cal D}_{c d}\;
\frac{\pi \alpha_s^2(p_\perp^2)}{\hat s^2}\;\,
\vert \overline{\cal M}_{a b \rightarrow c d}\vert ^2
\;\;,
\end{equation}
where the degeneracy factors ${\cal D}_{ab} = (1+ \delta_{ab})^{-1}$
accounts for the identical particle effect in the initial state 
if $a$ and $b$ are truly indistinguishable,
and correspondingly ${\cal D}_{cd}$ is the statistical factor for 
the final state.  The squared matrix elements for the diagrams in Fig. 9 
are \cite{fields}:
\begin{eqnarray}
\vert \bar{\cal M}_{q_i q_j \rightarrow q_i q_j}\vert^2 & = &
\frac{4}{9} \left(\frac{\hat{s}^2 + \hat{u}^2}{\hat{t}^2} \right) \;+\;
\delta_{i j} \; \left[ \frac{4}{9} \left(\frac{\hat{s}^2 + \hat{t}^2}{\hat{u}^2} \right) \;-\;
\frac{8}{27} \left(\frac{\hat{s}^2}{\hat{u} \hat{t}} \right)
\right]
\nonumber
\\
\vert \bar{\cal M}_{q_i \bar q_j \rightarrow q_i \bar q_j}\vert^2 & = &
\frac{4}{9} \left(\frac{\hat{s}^2 + \hat{u}^2}{\hat{t}^2} \right) \;+\;
\delta_{i j} \; \left[ \frac{4}{9} \left(\frac{\hat{t}^2 + \hat{u}^2}{\hat{s}^2} \right) \;-\;
\frac{8}{27} \left(\frac{\hat{u}^2}{\hat{s} \hat{t}} \right)
\right]
\nonumber
\\
\vert \bar{\cal M}_{q_i g \rightarrow q_i g}\vert^2 & = &
- \frac{4}{9} \left(\frac{\hat{u}^2 + \hat{s}^2}{\hat{u} \hat{s}} \right) \;+\;
 \left(\frac{\hat{u}^2 + \hat{s}^2}{\hat{t}^2} \right)
\nonumber
\\
\vert \bar{\cal M}_{g g \rightarrow g g}\vert^2 & = &
\frac{9}{2} \left(3 \;-\;\frac{\hat{u} \hat{t}}{\hat{s}^2} \;-\;
\frac{\hat{u} \hat{s}}{\hat{t}^2} \;-\;
\frac{\hat{s} \hat{t}}{\hat{u}^2} \right)
\;\;,
\end{eqnarray}
The variables $\hat s$, $\hat t$, $\hat u$ are the kinematic invariants of the parton-parton scattering
with
$\hat s + \hat t + \hat u = 0$, and $p_\perp^2 = \hat t \hat u / \hat s$
for massless particles.  For massive quarks the corresponding scattering 
matrix-elements can be found in Ref. \cite{combridge,svetitsky}.
\smallskip

\noindent
{\bf 3.4 The evolution equations for the parton shower functions}
\smallskip

Adding all the interaction rates $R^{(1\rightarrow 2)}$, 
$R^{(2\rightarrow 1)}$, and $R^{(2 \rightarrow 2)}$, we  obtain the 
following set of evolution equations for the parton densities $Q_i$ of 
quarks, $\bar Q_i$ of antiquarks, and $G$ of gluons, respectively:
\begin{eqnarray}
\frac{\partial}{\partial t} \, Q_i(x, \tau,t) &=&
- \;\left[
\;\hat A  \; + \; 
\hat A^\prime[\hat G] \;+\; \hat B^\prime[\hat{\bar{Q}}_i]  \right] \; Q_i 
\,\;+\;\,\left[  \; \hat B \;+\; 
\hat S_{q g}[\hat Q_i] \right]\; G
\nonumber
\\
& &\;\;+\;
\left[
\sum_{j=1}^f 
\hat S_{q q}[\hat Q_i] \right] \; Q_j \;+\; 
\left[
\sum_{j=1}^f 
\hat S_{q \bar q}[\hat Q_i] \right]\; \bar Q_j 
\label{eq}
\\
\,\frac{\partial}{\partial t} \, \bar Q_i(x, \tau, t) &=&
- \;\left[
\;\hat A  \; + \;
\hat A^\prime[\hat G] \;+\; \hat B^\prime[\hat{Q}_i]  \right] \; \bar Q_i 
\,\;+\;\,\left[  \; \hat B \;+\; 
\hat S_{q g}[\hat{\bar{Q}}_i] \right]\; G
\nonumber
\\
& &\;\;+\;
\left[
\sum_{j=1}^f 
\hat S_{q \bar q}[\hat{\bar{Q}}_i]\right] \; Q_j 
\;+\; 
\left[
\sum_{j=1}^f 
\hat S_{\bar q \bar q}[\hat{\bar{Q}}_i]\right] \; \bar Q_j 
\label{eqq}
\\
\,\frac{\partial}{\partial t} \, G(x, \tau, t) &=&
- \;\left[
\;\hat C \;+\;\hat D  \; + \; 
\hat C^\prime[\hat G]
\;+\; \sum_{j=1}^f \left\{ D^\prime[\hat Q_j] \,+\,\hat D^\prime [\hat{\bar{Q}}_j] \right\}
\;-\;
\hat S_{g g}[\hat G] \right]\; G
\nonumber
\\
& &
\;+\; \sum_{j=1}^f \left[\frac{}{}
\; \hat E \;+\; 
\hat E^\prime [\hat{\bar{Q}}_j]  \;+\;
\hat S_{g q}[\hat G] \right] \; Q_j 
\;+\; \sum_{j=1}^f \left[\frac{}{}
\; \hat E \;+\;
\hat E^\prime [\hat Q_j]  \;+\;
\hat S_{g \bar q}[\hat G] \right]\; \bar Q_j 
\;\;.
\label{eg}
\end{eqnarray}
Note that this set of equations actually describes the time evolution of
the parton (longitudinal) momentum distributions $Q = x q$, $\bar Q = x 
\bar q$ and $G = x g$, rather than of the parton number densities $q$, 
$\bar q$ and $g$. Similarly, one can write down in the full evolution 
equations for the first moments in $k_\perp^2$,
\begin{eqnarray}
\Pi_{q_i}(x, \tau, t) &=& 
\int dk_\perp^2 \;k_\perp^2 \;x\,q_i(x, k_\perp^2, \tau, t)
\nonumber \\
\Pi_g(x, \tau, t) &=&
\int dk_\perp^2 
\;k_\perp^2\;x \,g(x, k_\perp^2, \tau, t)
\;\;.
\end{eqnarray}
in straightforward generalization of eq. (\ref{peveq}).

Finally we stress that the evolution equations (\ref{eq})-(\ref{eg}) can
immediately be generalized to treat the cascading partons and the
partons of the nuclear medium on the same footing, by dropping our
bookkeeping distinction among those two particle sources. 
This would then describe a dynamically coupled system in which the parton
cascade evolution feeds back on the nuclear parton distribution.
The only differences to eqs. (\ref{eq})-(\ref{eg}) are that the nuclear
parton densities also become time-dependent, i.e. $\hat a(x) \rightarrow \hat a(x,t)$,
where $a=q_i,\bar q_i, g$, and in the gain term of $gg \rightarrow g$ fusion, 
as well as in the $gg$ scattering
rates an additional factor of $1/2$ would be needed, because the two interacting
gluons, a cascade parton and a nuclear parton, cannot be distinguished anymore.
However, the response of the nuclear density to the penetrating parton shower
is naturally delayed and only locally effective, so that the parton cascade
to a good approximation can be viewed as being unaffected by this time variation
of the nuclear medium. Of course this approximation does not apply anymore 
when there are multiple cascades evolving simultaneously close to each other 
in phase-space, as e.g. in a nucleus-nucleus collision.
In such a case the full space-time history of both the cascading partons as well
as the nuclear partons needs certainly to be taken into account.
\bigskip
%\newpage

\noindent {\bf 4. SUMMARY}
\medskip

To summarize the essence of this work, let us list what we believe are 
the most important points:
\itemitem{(i)}
We have derived integro-differential equations for the evolution of a 
parton cascade in an infinite, homogeneous nuclear medium, describing the 
parton distributions in terms of the Bjorken variable $x$ and virtuality 
$Q^2$ (or ``age'' $\tau$).
\itemitem{(ii)}
The Lorentz invariant evolution equations have the character of transport 
equations in momentum space, familiar from non-equilibrium kinetic theory, 
however, they include effects of off-shell propagation in addition
to collision terms.
\itemitem{(iii)}
In the absence of a medium the equations reduce to the 
Altarelli-Parisi-Lipatov equations for the $Q^2$-evolution of the parton 
number densities in vacuum.
\itemitem{(iv)}
Possible immediate applications of the evolution equations are, for 
example, to the fragmentation of partons in heavy nuclei, and to hard 
QCD probes of a quark-gluon plasma.
\itemitem{(v)}
The equations can be easily generalized to provide a description of 
parton transport in ultra-relativistic heavy ion collisions by
treating the shower partons and the nuclear partons on the same footing.
\itemitem{(vi)}
Since our derivation was partially based on heuristic arguments, it would be 
desirable to obtain a formal justification from fundamental principles of
quantum field theory by Green's function methods.

We hope to address these issues in future publications.
\bigskip

\section*{Acknowledgements}
\vspace{-2mm}
We thank J. I. Kapusta and E. V. Shuryak for interesting discussions.
Much of this work was done at the Institute for Theoretical Physics (ITP)
at the University of California, Santa Barbara, during the program
``Strong interactions at finite temperatures''.
One of us (K.G.)  acknowledges fruitful conversations with A. H. Mueller at Columbia University. 
This work was supported in part by the 
National Science Foundation under Grant No. PHY89-04035 and by the U.S.
Department of Energy (grant numbers DE-FG05-90ER40592 and DE-FG02-87ER-40328).
\bigskip

%\newpage

\newpage

{\bf FIGURE CAPTIONS}
\bigskip

\noindent {\bf Figure 1:}  
Visualization of
a parton cascade evolving in the quark-gluon matter of a nucleus.
A primary parton
that has been produced at some point of time $t_0$ with a time-like virtuality $Q_02$
initiates a shower of secondary partons 
by gluon bremsstrahlung and multiple interactions with the nuclear medium.
\medskip

\noindent {\bf Figure 2:}  
Illustration of a $pA$ collision in the CM$_{NN}$ frame. The incident proton
sees the nucleus as a Lorentz contracted ensemble of virtual
gluons and quarks \cite{pcm0,pcmpa}. Similarly, due to the symmetry of the
reference frame, the proton itself appears to the nucleus
as an incident distribution of individual partons, smeared out along the longitudinal
direction.
The longitudinal momenta of the partons from the proton are taken as $p_z = x P$ and the 
nuclear partons have $p_z^\prime = - x^\prime P$.
\medskip

\noindent {\bf Figure 3:}  
a) Diagram of
the first order transition amplitude $w^{(0)}(t)$ for a
an initial gluon state $\vert i\rangle$ to be converted into a state $\vert f \rangle$ by 
means of an external interaction $V_0(t_0)$ with the final gluon being on mass shell.

b) Diagram of the amplitude $w^{(1)}(t)$ for the second order process where 
an intermediate virtual state $\vert a \rangle$
is produced 
that subsequently decays into the  2-gluon final state $\vert f \rangle=\vert bc\rangle$
at time $t_1$ according to the decay matrix element  $V_1(E_a)$.
\medskip

\noindent {\bf Figure 4:}  
a) Evolution of a gluon cascade in terms of successive branchings
the LLA with many intermediate
virtual states. The cascade is initiated by a primary gluon with virtuality $Q_0^2$ at
time $t_0$ and develops downwards
with strongly ordered decreasing virtualities $Q_i^2 \gg Q_{i+1}^2$ from $Q_0^2$ to  $\mu_0^2$,
corresponding to the time evolution from $t_0$ to $t_f$ with
$t_i \ll t_{i+1}$ and $t_f(\mu_0^2)$.

b) Corresponding longitudinal and lateral spread of the multigluon wavefunction as time
progresses. The diffusion in $r_\perp$-direction results in a linear (with time) growing
cross-section of the cascade when evolving in  a medium, as inside a nucleus.
\medskip

\noindent {\bf Figure 5:}  
Graphical representation of the time-evolution of the gluon distribution 
in terms of an infinite sum of contributions
involving $n$ successive branchings, $g(x,t) = \sum_n g^{(n)}(x,t)$.
\medskip

\noindent {\bf Figure 6:}  
Connection between the ``age'' $\tau$ of a cascade parton and the time $t$. In vacuum
where the particle evolves (in the LLA) solely by successive branchings with decreasing virtualities $Q^2$,
the age of a parton is
measured in time via $t= x P/Q^2$.
In medium, however, scattering and fusion processes modify the age by rejuvenating the parton
in each interaction with the medium. That is, a parton ages slower, because it is shifted 
back to larger virtuality corresponding to the
transverse momentum squared exchanged in the scattering, or to the invariant mass of the compound
state in fusions, respectively.
\medskip

\noindent {\bf Figure 7:}  
Feynman diagrams associated with the gain and loss of a) quarks (antiquarks) and
b) of gluons due to elementary branching processes. The parton with momentum
fraction $x$ is the ``observed'' particle.
\medskip

\noindent {\bf Figure 8:}  
Feynman diagrams contributing to the gain and loss of a) quarks (antiquarks) and
b) of gluons by fusions with nuclear partons. The parton of interest is the one with momentum
fraction $x$ and the nuclear parton from the nucleus structure function has momentum
fraction $x_2$.
\medskip

\noindent {\bf Figure 9:}  
Diagrams of
elementary scattering processes that increase of the number of a) quarks (antiquarks) and
b) of gluons with momentum fraction $x$ due to the liberation of a virtual nuclear parton (labelled by a ``hat'')
out of the wavefunction of the nucleus by an interaction with a cascade parton.
\medskip

\vfill
\end{document}